\documentclass[apj,numberedappendix]{emulateapj}
\usepackage{amsmath}
\usepackage{hyperref}

\usepackage[english]{babel}
\usepackage{graphicx}
\usepackage{latexsym}
\usepackage{amsmath,amssymb}
\usepackage{natbib}
\usepackage{dcolumn}
\usepackage{bm}
\usepackage{calrsfs}
\usepackage{ulem}
\usepackage[usenames]{color}
\usepackage{xspace}

\newcommand{\nn}{\mbox{} \nonumber \\ \mbox{} }
\newcommand{\ba}{\begin{eqnarray}}
\newcommand{\ea}{\end{eqnarray}}
\newcommand{\om}{\omega}

\newcommand{\Bf}{{magnetic field}}

\newcommand{\NS}{neutron star}

\newcommand{\Lf}{{Lorentz factor}}

\newcommand\eg{\textit{e.g.}}
\newcommand\cf{\textit{cf.\ }}

\def\be{\begin{equation}}
\def\ee{\end{equation}}

\begin{document}

\title{Dynamics  and emission of  wind-powered  afterglows of gamma-ray bursts: flares, plateaus and steep decays}


\author{Maxim V.~Barkov$^{1}$, Yonggang Luo$^{2}$ and 
Maxim Lyutikov$^{2}$\\
 $^1$Institute of Astronomy, Russian Academy of Sciences, Moscow, 119017 Russia\footnote{barkov@inasan.ru }\\
 $^2$Department of Physics and Astronomy, Purdue University, 525 Northwestern Avenue, West Lafayette, IN 47907-2036, USA   }

\begin{abstract}
We develop a model of early X-ray afterglows of gamma-ray bursts originating from the reverse shock  (RS) propagating through ultra-relativistic, highly  magnetized  pulsar-like  winds produced by long-lasting  central engines.  
We first perform fluid and MHD numerical simulations  of relativistic double explosions. We demonstrate that even for  constant properties of the wind  a variety of temporal behaviors can be produced, depending on the
 energy of the initial explosion and    the wind power,  the delay time for the switch-on of the wind, and  magnetization of the wind.     
  X-ray emission of the highly magnetized RS occurs in the fast cooling regime - this  ensures high radiative efficiency  and allows fast intensity variations.  We demonstrate that:  (i) RS emission naturally produces light curves showing power-law temporal evolution with various temporal indices;
(ii) mild wind power, of the order of  $\sim 10^{46}$ erg s$^{-1}$ (equivalent isotropic), can reproduce the  afterglows' plateau phase; 
(iii)   termination of the  wind can  produce sudden steep decays;  (iv) short-duration afterglow flares are due to mild variations in the wind luminosity,  with small total injected energy.
\end{abstract}


\maketitle  

\section{Introduction}
\label{s:in}

Gamma-ray bursts (GRBs) are produced in relativistic explosions \citep{1986ApJ...308L..43P,2004RvMP...76.1143P} that 
generate two shocks: forward shock and reversed shock. The standard fireball model \citep{1992MNRAS.258P..41R,1995ApJ...455L.143S,1999PhR...314..575P,2006RPPh...69.2259M} postulates that 
the prompt emission is produced by internal  dissipative  processes  within the flow: collisions of matter-dominated shells, \cite{1999PhR...314..575P}, or reconnection events \citep{2006NJPh....8..119L}). The   afterglows, according to the fireball model, are generated in 
the external relativistic blast wave.

Since emission from the forward shock depends on ``integrated properties'' (total injected energy and total swept-up mass), the corresponding light curves were expected to be fairly smooth. In contrast, 
observations  show the  presence of unexpected  features like  flares  and  light curves plateaus 
\citep{2006ApJ...642..389N,2006ApJ...647.1213O,2013FrPhy...8..661G,2016ApJ...829....7L,2007MNRAS.377.1638D,2010MNRAS.406.2113C,2018arXiv180410441M}, abrupt endings of the plateau phases \citep{2007ApJ...665..599T}, 
fast optical variability (\eg\ GRB021004 and most notoriously GRB080916C), missing  \citep{2016MNRAS.462.1111D} and chromatic \citep{Panaitescu2007,Racusin09}  
jet breaks, missing reverse shocks \citep{Gomboc:2009}). These phenomena  are hard to explain within  the standard fireball model that postulates  that the early $X$-ray are produced in the forward shock, as argued by  \citep{lyutikov_09,2010ApJ...720.1513K,2017ApJ...835..206L}. 

          The origin of sudden drops in afterglow light curves  is especially  mysterious. As an example,
  GRB 070110 starts with a normal prompt emission, followed 
by an early decay phase until approximately  100 seconds,  and a plateau until $\sim 10^4$ s.
 At  about $2 \times 10^4$ seconds, the light 
curve of the afterglow of GRB 070110 drops suddenly with a temporal  slope $> 7$ \citep{2007GCN..6008....1S,2007GCN..6014....1K,2007GCNR...26....2K,2007ApJ...665..599T}.

Observations of early afterglows in long Gamma Ray Bursts (GRBs), at times $\leq$ 1 day, require a presence of long-lasting active central  engine.
{Previously, some of the related phenomenology was  attributed to long lasting central engine (see \S \ref{modelss} for a more detailed discussion of various models of  long-lasting central engine).  A number of authors discussed  long-lasting engine that produces colliding  shells, in analogy with the fireball model for the  prompt emission 
\citep{1994ApJ...430L..93R,2006MNRAS.366.1357P,2007ApJ...665L..93U,2010MNRAS.401.1644B,2011MNRAS.417.2161B}. 
The problem with this explanation is that energizing the forward shock requires a lot of energy: the total energy in the blast needs to increase linearly with time, hence putting exceptional demands 
on the efficiency of prompt emission \citep{2006MNRAS.366.1357P,2007MNRAS.380..270O,2009MNRAS.392..153D}. 
In addition, to produce afterglow flares in the forward shock the total energy in the explosion needs to roughly double each time: hence the total energy grows exponentially for bursts with multiple flares.}

As an alternative, \cite{2017ApJ...835..206L}  developed  a  model of early GRB afterglows  with  dominant $X$-ray contribution from the reverse shock (RS) 
propagating in  highly relativistic (Lorentz factor $\gamma_w \sim 10^4-10^6$) magnetized wind of a  long-lasting central engine; we will refer to this types of model as "a pulsar paradigm", stressing similarities to physics of pulsar winds.

Pulsar wind Nebulae (PWNe) are efficient in converting spindown energy of the central objets, coming out in a form of the wind, into high energy radiation, reaching efficiencies of tens of percent \citep[\eg][]{1984ApJ...283..710K,2008AIPC..983..171K}. This efficiency is much higher that what would have been expected from simple sigma-scaling of dissipation at relativistic shocks \citep{1984ApJ...283..694K}.
Effects of magnetic dissipation contribute to higher efficiency \citep{2011ApJ...741...39S,2014MNRAS.438..278P}.

  \cite{2017ApJ...835..206L}  adopted the pulsar wind model to the case of preceding expanding GRB shock.
 The model  reproduces, in a fairly natural way,  the overall trends and yet allows for variations in the temporal and spectral evolution of early 
optical and $X$-ray afterglows.  The high energy and the optical synchrotron emission from the RS particles occurs in the fast cooling regime;  
the resulting synchrotron power $L_s$ is a large fraction of the wind luminosity (high-sigma termination shocks propagate faster through the wind, boosting the efficiency.)

 Thus, plateaus - parts of afterglow light curves that show slowly decreasing spectral power  -  
are a natural consequence of the RS emission.  Contribution from the forward shock (FS) is negligible in the $X$-rays, but  in the optical both 
FS and RS contribute similarly \citep[but see, e.g., ][]{2017ApJ...835..248W,2018MNRAS.480.4060W,2019NatCo..10.1504I}:  the FS optical emission is in the slow cooling regime, producing smooth components, while the RS optical 
emission is in the fast cooling regime, and thus can both produce optical plateaus  and account for  fast optical variability  correlated 
with the  $X$-rays, e.g., due to   changes in the wind properties. The later phases of pulsar wind interaction with super nova remnant discussed by  \cite{2018ApJ...860...59K}.

The goal of the present work is two-fold. First,  we perform a number of numerical simulations for the propagation of a highly relativistic  magnetized wind that follows a  relativistic shock wave. 
Previously, this problem was considered analytically by \cite{2017PhFl...29d7101L}.
Second, we perform radiative calculations of the early  X-ray afterglow emission coming from 
the  ultra-relativistic RS of a long-living central engine. We demonstrate that this paradigm allows us to resolve the problems of plateaus, sudden intensity drops, and flares. Qualitatively,   at early times, a large fraction of the wind power is radiated: this explains the plateaus. If the wind terminates, so that the  emission from RS  ceases  instantaneously,
 this will lead to a sharp decrease in observed flux (since particles are cooling fast). Finally, variations of the wind intensity can  produce  flares that bear resemblance to the  ones  observed in GRBs.

We argue in this paper that abrupt declines in afterglow curves can be explained if emission originates in the ultra-relativistic and highly magnetized   reverse shock of a long-lasting engine.
\cite{2017ApJ...835..206L} \citep[see also][]{2017PhFl...29d7101L} developed a model of early GRB afterglows with dominant X-ray contribution from the highly magnetized ultra-relativistic  reverse shock (RS), an analog of the pulsar wind termination shock.
The critical point is that emission from the RS in highly magnetized pulsar-like wind  occurs in the fast cooling regime. Thus it  reflects {\it instantaneous} wind power, not accumulated mass/energy, as in the case of the forward shock. Thus, it is more natural to produce fast variation in the  highly magnetized RS.

\section{Models of long-lasting winds in GRBs}
\label{modelss}

The model of  \cite{2017ApJ...835..206L}, explored in more details here,   differs qualitatively  from a number of previous works that advocated a long lasting central engine in GRBs. Previous works can be divided into two categories. First  type of models involves modifying the properties of the forward shock (FS) \citep[\eg\ re-energizing of the FS by the long-lasting wind in an attempt to produce flares][]{1998ApJ...496L...1R,1998A&A...333L..87D,1998ApJ...503..314P,2004ApJ...606.1000D}.
The second type of models assume  a 
long lasting central engine that produces  mildly relativistic matter-dominated  winds \citep{1999ApJ...517L.109S,2007MNRAS.381..732G,2007ApJ...665L..93U,2009MNRAS.397.1153K,2012ApJ...761..147U,2017MNRAS.472L..94H}. In these types of mode the   emission is produced  in a way similar to the internal shock model for the prompt  emission (that is, collision of baryon-dominated shells, amplification of \Bf\ and particle acceleration). 

The FS-based models encounter a number of fundamental problems \citep{lyutikov_09,2007ApJ...665L..93U} 
\cite[Though see][]{2010MNRAS.402..705L,2010MNRAS.409..531R,2016ApJ...825...48R,2017A&A...605A..60B,2013MNRAS.430.1061R,2014MNRAS.442.3495V,2020arXiv200300927K,2020arXiv201006234W}.  The key problem is that the properties of the forward shock are ``cumulative'', in  a sense that its dynamics depend on the {\it total}  swept-up mass and injected energy, which is impossible to change on a short time scale. For example, to produce a flare within the FS model, the total energy  of the shock should increase substantially (\eg, by a factor of two). To produce another flare, even more energy need to be injected, leading to the exponentially increasing total energy with each flare. 

Most importantly, the FS-based models cannot produce  abrupt steep decays. Such sharp drops require (at the  least) that  the  emission from the forward shock (FS) switches off instantaneously.  This is  impossible.  First,  the microphysics of shock acceleration is not expected to change rapidly (at least we have no arguments why it should).

Second, the variations of  hydrodynamic properties of the FS, as they translate to radiation,  are also expected to produce smooth variations \citep{2013ApJ...773....2G}.
As an example,   consider a relativistic  shock that breaks out from a denser medium (density $n_1$) into the less dense one (density $n_2\ll  n_1$). In the standard fireball model total  synchrotron power $P_s$ per unit area of the shock   scale as \citep{2004RvMP...76.1143P}
\ba &&
P_s \propto n \Gamma^2 \gamma^{\prime 2} B^{\prime 2} \propto n^2 \Gamma^6
\nn &&
\gamma^{\prime} \propto \Gamma
\nn && 
B^{\prime}  \propto \Gamma \sqrt{n}
\ea
where $\Gamma$ is the \Lf\ of the shock,  $\gamma^{\prime} $ is the \Lf\ of accelerated particles.

 Importantly, if  a shock breaks out from a dense medium into the rarefied one, with $n_2\ll  n_1$, it {\it accelerates} to approximately $\Gamma_2 \approx \Gamma_1^2$, as the post-shock internal energy in the first medium is converted into bulk motion  \citep {1971PhRvD...3..858J,2010PhRvE..82e6305L}. 
Thus a change in power and peak frequency scale as
\be 
\frac{P_{s,2}}{P_{s,1}} ={\Gamma_1^6} \left(  \frac{n_2}{n_1}\right)^2
\ee
Thus, even though we assumed $n_2 \ll n_1$, the synchrotron emissivity in the less dense medium is largely  compensated by the increase of the \Lf. 
 Since the expected \Lf\ at the  time of sharp drops is $ \Gamma_1 \sim$ few tens, suppression of emission from the forward shock requires the unrealistically large decrease of  density. 
 
 {
 \cite{2020ApJ...893...88O} discussed appearance of plateaus from an off-axis jet (so that a more energetic part of the FS becomes visible and effectively boosts the observed flux. We expect though that at  observer times of few$\times 10^4$ seconds the X-ray emitting particles  in the FS are  in the slow cooling regime,  liming how short time scales in the observed  emission light curves can be produced.
}

The model of  \cite{2017ApJ...835..206L},  and the present investigation, is more aligned with the previously discussed  emission from the RS
\citep{1999ApJ...517L.109S,2007MNRAS.381..732G,2007ApJ...665L..93U,2012ApJ...761..147U,2017MNRAS.472L..94H}.
 But the present model is qualitatively different: emission properties here  are parametrized  within the pulsar wind paradigm of \cite{1984ApJ...283..710K}, not the fireball model \citep[\eg][]{2004RvMP...76.1143P}. Qualitatively, the advantage of the  present model of the highly magnetized/highly relativistic  RS emission over the fireball adaptation to the RS case are similar to the prompt emission: high magnetized relativistic flows can be more efficient in converting the energy of the explosion to radiation, as they do not ``lose'' energy on the bulk motion of non-emitting ions \citep{lb03,2006NJPh....8..119L}.

  The pulsar wind paradigm of \cite{1984ApJ...283..710K} also has a very different prescription for particle acceleration and emission: it  relates the typical (minimum) \Lf\ of the accelerated particles to the \Lf\ of the pre-shock wind $\gamma _{min} \sim  \gamma_w$ with $\gamma_w \sim 10^4-10^6$), while the magnetic field in the emission region follows the shock compression relations. In contrast, the fireball model parametrizes both the   \Lf\ of the accelerated particles and the (shock-amplified) \Bf\ to the  upstream  properties of the baryon-dominated energy flow (\eg, $\gamma _{min} \sim \epsilon_e (m_p/m_e) \gamma_w$ with $\gamma_w \sim 10^2$). The resulting emission properties are qualitatively different.

 \section{Relativistic  double explosion} 
\label{s:expect}

\subsection{Triple shock structure}

Consider relativistic point explosion of energy $E_1$ in a medium with constant density $\rho_{ex}=m_p n_{ex}$,  followed by a wind with  constant luminosity $L_w$ \citep[both $E_1$ and $L_w$ are isotropic equivalent values]{2017PhFl...29d7101L}. 
The initial explosion generates a Blandford-McKee  forward  shock wave  (BMFS)  \cite{bm76}
\ba &&
\Gamma_1 = \sqrt{ \frac{17}{8 \pi }}   
\sqrt{ \frac{E_1}{\rho_{ex} c^5}}  t^{-3/2}
\nn &&
p_1= \frac{2}{3} \rho_{ex} c^2 \Gamma_1 ^2 f_1(\chi)
\nn &&
\gamma^2_1 = \frac{1}{2}\Gamma_1 ^2 g_1(\chi)
\nn &&
n_1= 2 n_{ex} \Gamma_1 n_1(\chi)
\nn &&
f_1(\chi)=\chi^{-17/12} 
\nn && 
g_1(\chi) =1/\chi
\nn &&
n_1(\chi)= \chi^{-5/4}\
\nn &&
\chi = \left[1+2(m+1)\Gamma^2\right]\left(1-r/t\right)
\label{eq:bmc}
\ea
Subscript $ex$ indicates the properties in the surrounding medium;
subscript $1$ indicates that quantities are measured behind the leading BMFS, hence between the two forward  shocks;  The  Lorentz factor $\Gamma$  depends on time as $\Gamma^2\propto t^{-m}$, $m=3$.

We assume that the initial GRB explosion leaves behind an active  remnant - a black hole or (fast rotating) \NS. The remnant  produces a long-lasting pulsar-like  wind, either using the rotational energy of the newly born \NS\ \citep{Usov92,2007MNRAS.382.1029K}, accretion of the pre-explosion envelope onto the BH \citep{CannizzoGehrels}, or if the black hole can keep its magnetic flux for sufficiently long time
\citep{2009MNRAS.397.1153K,2010MNRAS.401.1644B,2011PhRvD..83l4035L,2011PhRvD..84h4019L}. 

\begin{figure}
 \centering
\includegraphics[width=.9\columnwidth]{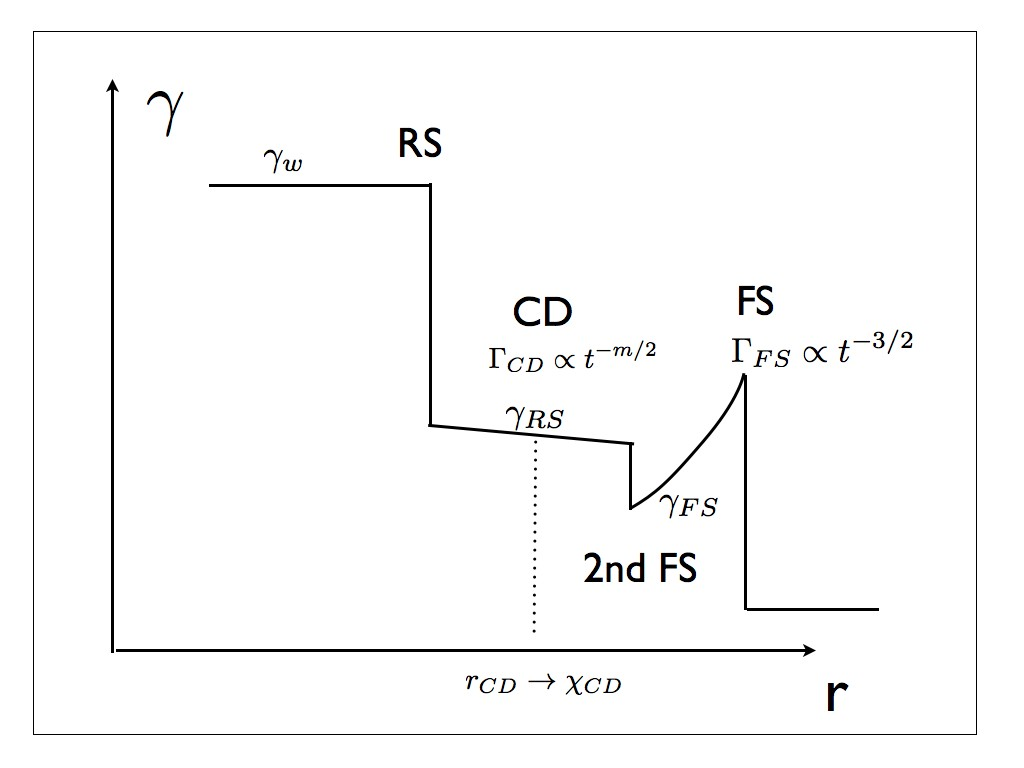}
  \caption{Velocity structure of the triple-shock configuration.  Leading is the FS that generates a  self-similar post-shock velocity and pressure profiles. A fast wind with \Lf\ $\gamma_w$ is terminated at the reverse shock (RS); the post-RS flow connects through the  contact discontinuity (CD, dotted line)  to the second shock driven in the already shock media. The CD is located at $r_{\rm CD}$, corresponding to $\chi_{\rm CD}$.  The RS and the second forward shock (2nd FS) are located close to $\chi_{\rm CD}$ \protect\citep{2017ApJ...835..206L}.
}
 \label{Shock-structure-gamma}
\end{figure}

One expects that the central engine produces very fast and light wind that will start interacting with the slower, but still relativistically expanding, ejecta.
As the highly relativistic  wind from the long-lasting engine interacts with the initial explosion, 
it launches a second forward shock in the medium already shocked by the primary blast wave. At the same time   the reverse shock forms in the wind; the two shocks are separated by the contact discontinuity (CD), Figure \ref{Shock-structure-gamma}.

First, we assume that  external density is constant, while the  wind is  magnetized  with constant luminosity (variations in wind luminosity are explored in \S  \ref{rad})
\begin{equation}
L_{w}=4\pi\gamma_{w}^{2}\left(\rho_wc^{2}+\frac{B_{w}^{2}}{4\pi}\right)r^{2}c
\end{equation}
where $\rho_{w}$ and $B_{w}$ are density and \Bf\ measured  in the wind rest frame. Thus
\begin{equation}
B_{w}=\sqrt[]{\frac{\sigma_w}{1+\sigma_w}} \ \sqrt[]{\frac{L_{w}}{c}}\frac{1}{r\gamma_{w}}
\label{lbw}
\end{equation}
where
\begin{equation}
\sigma_w=\frac{B^{2}_{w}}{4\pi \rho_wc^{2}}.
\end{equation}
is the wind magnetization parameter \citep{1984ApJ...283..694K}. In our ``pulsar wind'' paradigm, we assume that the mass loading of the wind is very small, while the 
the  wind is  assumed to be very fast,  with  $\gamma_w \gg \Gamma_{FS},\, \Gamma_{\rm CD}$.

\subsection{Analytical expectations: self-similar stages} 

Generally, the structure of the flows in double explosions is  non-self-similar \citep{2017PhFl...29d7101L}. First, with time the second forward shock approaches the initial  forward shock (FS);   for sufficiently powerful winds the second FS may catch up with the primary FS. The presence of this special time violates the assumption of self-similarity.
We can estimate the catch-up time by noticing that the power deposited by the wind in the shocked medium scales as $L_w/\Gamma_{\rm CD}^2$. 
 Thus,  
in coordinate time the wind deposits energy similar to the initial explosion at time when $\Gamma_{\rm CD} \sim \Gamma_{FS}$,
\be
t_{eq}= \Gamma_{FS}^2 \frac{E_1}{L_w} \propto  \left(\frac{E_1^2}{c^5 \rho  L_w} \right)^{1/4}= 
2 \times 10^7 \; E_{1, 52}^{1/2} , L_{w, 46}^{-1/4} n^{-1/4} \mbox{ sec},
\label{teq}
\ee
almost a year in coordinate time.
At times $t \leq t_{eq}$ the second shock is approximately self-similar, the CD is located far downstream of the first shock;  and is moving with time in the self-similar coordinate $\chi$,  associated with the primary shock, towards the first shock. The motion of the first shock is unaffected by the wind at this stage.   
At times $t \geq t_{eq}$ the two shocks merge - the system then relaxes to a Blandford-McKee self-similar solution with energy supply.

In the numerical estimate in (\ref{teq}) we used the  wind power $L_{w}\sim 10^{46}$ erg s$^{-1}$  which at first glance may look too high. Indeed, the total energy budget for isotropic wind is then  $E_w\sim L_w  t_{eq}  \sim 10^{53}$~ergs, this value is much larger rotating energy of fast spinning NS $\sim 10^{52}$~erg. 
But recall that this is an isotropic equivalent power.
In the case of long GRBs, both  the initial explosion and the power of the long-lived central engine are collimated into small angle $\theta\sim 0.1$~rad \citep[\eg][]{2007MNRAS.382.1029K}.  After jet-break out the opening angle remains nearly constant. 
Thus, the true wind power can be estimated as $L_{w,true} \approx \theta^2 L_{w}/2 \sim 10^{44}$~erg/s and $E_w\sim 10^{51}$~ergs, 
which is an allowed energy budget of fast spinning NS.

 Secondly, the self-similarity may be violated at early times if there is  an effective delay time $t_d$ between the initial explosion and the start of the second wind. (This issues is also important in our implementation scheme, \S \ref{s:ss} - since we start simulation with energy injection at some finite distance from the primary shock this is equivalent to some effective time delay for the wind turn-on.)
 
Suppose that the secondary wind turns on  at  time $t_d$ after the initial one and the second shock/CD  is moving with the  
\Lf\ 
\be
\Gamma_{\rm CD} ^2 \propto (t-t_d)^{-m} 
\label{eq:lorcdcon}
\ee
Then, the location of the second shock at time $t$ is 
\be
R_{\rm CD} = (t-t_d) \left( 1-\frac{1}{2 \Gamma _{\rm CD}^2 (m+1)} \right)
\ee
The corresponding self-similar coordinate of the second shock in terms of the primary shock self-similar parameter $\chi$  is 
\be
\chi_{\rm CD} =\left(1 +8 \Gamma_1^2 \right) \left(1-\frac{R}{t}\right) 
\approx
\left( \frac{8 t_d}{t} + \frac{4}{(m+1) \Gamma_{\rm CD}^2} \right) \Gamma_1^2
\label{eq:td}
\ee

The effective time delay $t_d$ introduces additional  (beside the catch-up time (\ref{teq})) time scales in the problem. 
Thus, even within the limits of  expected self-similar motion, $t\ll t_{eq}$ the effective delay time $t_d$ violates the self-similarity assumption.
Still, depending  on whether the ratio $ t_d /( t\Gamma_{\rm CD}^2) $ is much larger or smaller than unity, we expect approximately self-similar behavior  \citep{2017PhFl...29d7101L,2017ApJ...835..206L}

For $t_d \geq t/(2 (m+1) \Gamma_{\rm CD}^2)$, the location of the CD in the self-similar coordinate associated with the first shock is 
\ba &&
\chi_{\rm CD} \approx
\frac{8 \gamma _1^2  {t_d}}{t}\propto t^{-4}
\label{eq:chit}
\\ &&
\Gamma_{\rm CD} =0.52 \frac{E_1^{5/48} {t_d}^{5/48} {L_w}^{1/4}}{c^{85/48} \rho ^{17/48} t^{11/12}}
   \label{eq:1112}
\ea
Alternatively, for $t_d \leq t/(2 (m+1) \Gamma_{\rm CD}^2)$,
\ba &&
\chi_{\rm CD}= 2.68 \left(\frac{E_1}{c^{5/2} \sqrt{\rho } t^2 \sqrt{L_w}}\right){}^{24/29}
\label{eq:2429}
\\ &&
\Gamma_{\rm CD}=0.50 \frac{E_1^{5/58} L_w^{6/29}}{c^{85/58} \rho ^{17/58} t^{39/58}}
\label{eq:3958}
\ea

Finally, if the second explosion is point-like with energy $E_2$, the \Lf\ of the second shock evolves according to \citep{2017PhFl...29d7101L}
\be
  \gamma_2 =\sqrt{\frac{71}{2}} \left(\frac{17}{\pi }\right)^{5/24}
\left(  \frac{{E_1}^5 {t_d}^5}{c^{85} (m_p n_{ex}) ^{17} }\right)^{1/24} \sqrt{E_2} t^{-7/3}
  \label{eq:Gamma22}
\ee
{ (this expression is applicable for $ t \leq \Gamma_1 ^2 t_d$, the time when the second shock catches with the primary shock.}

Relation (\ref{eq:1112}-\ref{eq:Gamma22}) indicate that depending on the particularities of the set-up, we expect somewhat different scalings for the propagation of the second shock (we are also often limited in  integration time to see a switch between different self-similar regimes). 

The point of the previous discussion is that   mild variations between the properties of double explosions (delay times, luminosity of the long lasting engine) are expected to produce a broad variety of behaviors, like    various power-law indices and  temporarily changing overall  behavior. This ability  of the model to accommodate a fairly wide range of behaviors with minimal numbers of parameters is  important in explaining highly temporally variable early afterglows, as we further explore  in this paper.

\section{Numerical simulations of relativistic double explosions}
\label{s:ss}

\subsection{Simulations' setup}
\begin{figure*}
\includegraphics[width=0.48\linewidth,angle=-0]{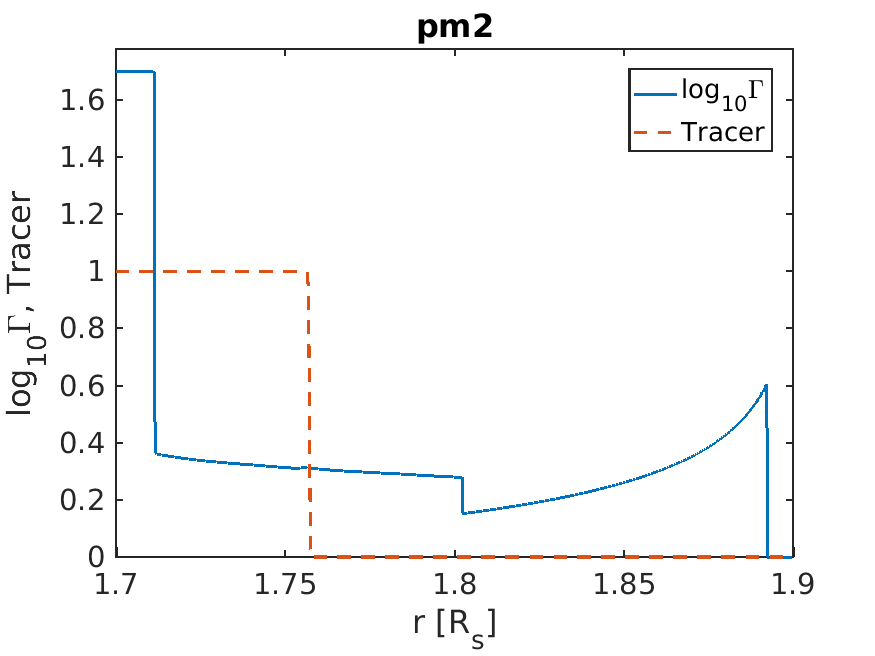}
\includegraphics[width=0.48\linewidth,angle=-0]{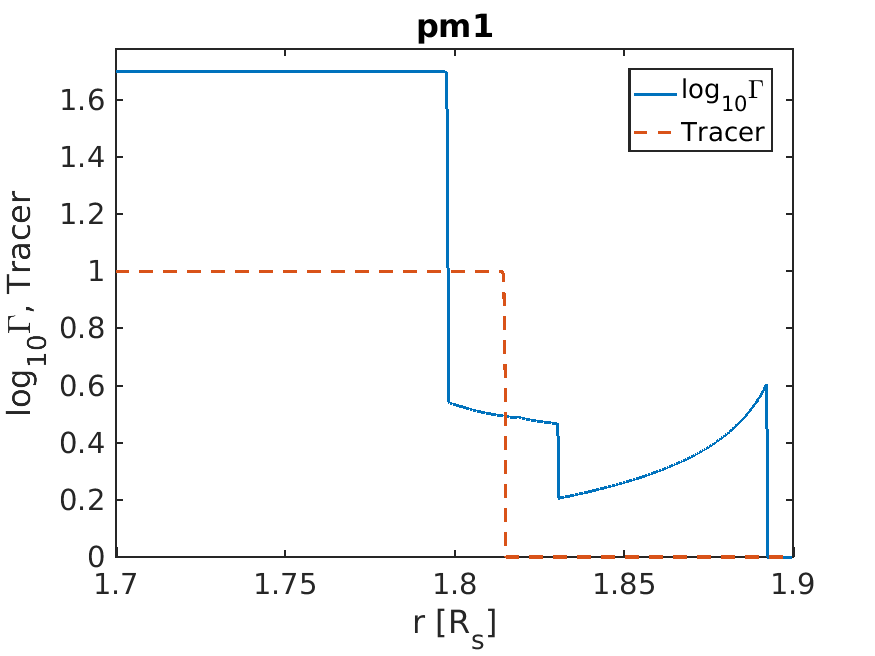}\\
\includegraphics[width=0.48\linewidth,angle=-0]{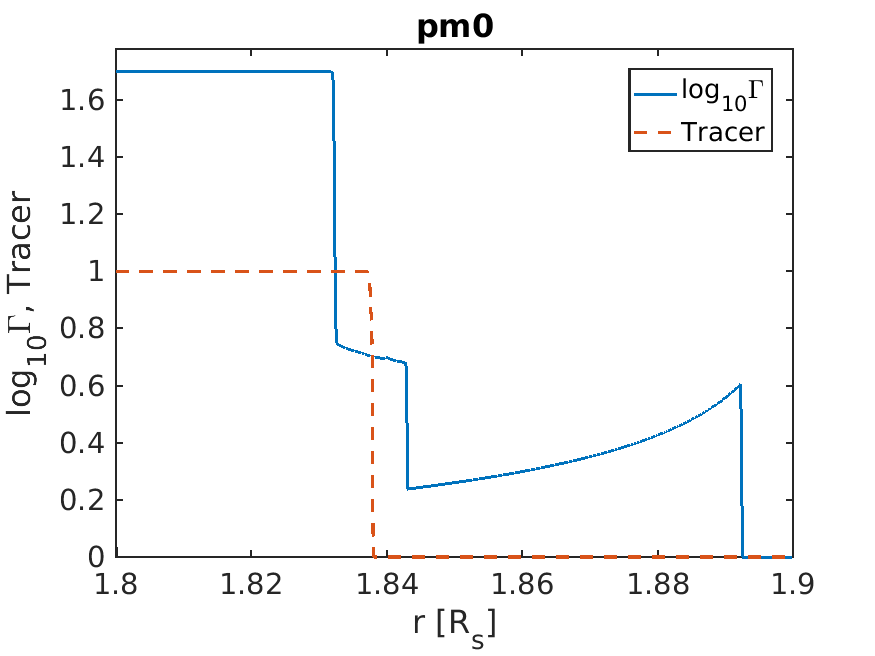}
\includegraphics[width=0.48\linewidth,angle=-0]{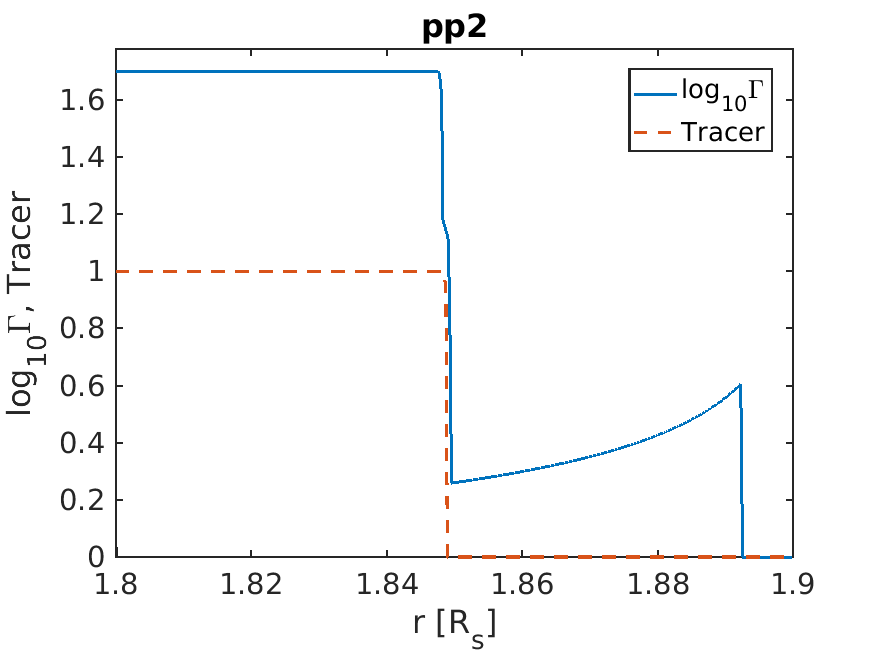}
\caption{Hydrodynamic simulations of the double explosion. Potted  are Lorentz factor and tracer distribution as a function of radius at the moment $t=1.9\; [r_s/c]$. The tracer distinguishes the wind from the shocked external medium. The parameters for each panel are encoded in the titles,  Table \protect \ref{tab:models}.}
\label{fig:LorTr}
\end{figure*}

\begin{figure*}
\includegraphics[width=0.48\linewidth,angle=-0]{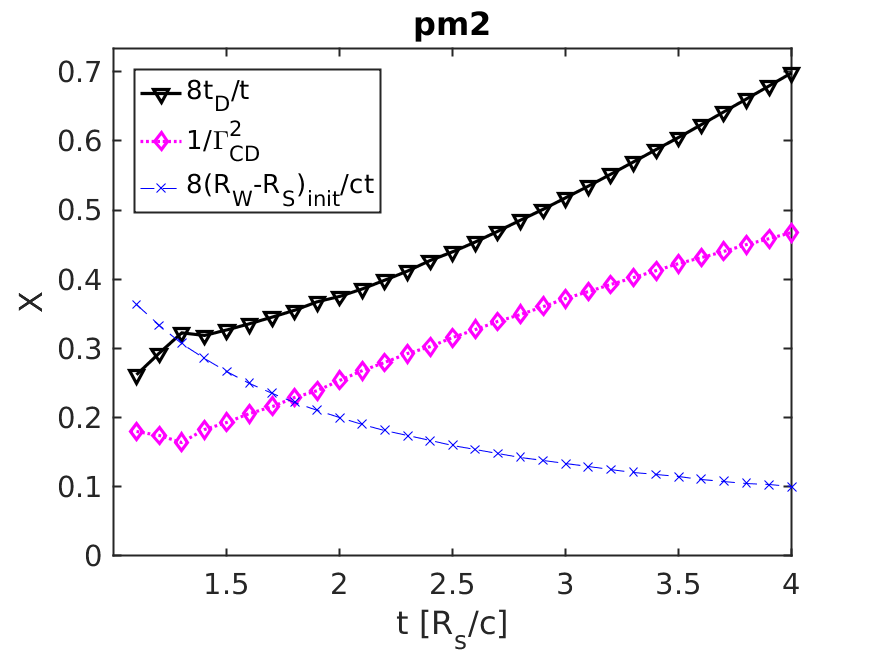}
\includegraphics[width=0.48\linewidth,angle=-0]{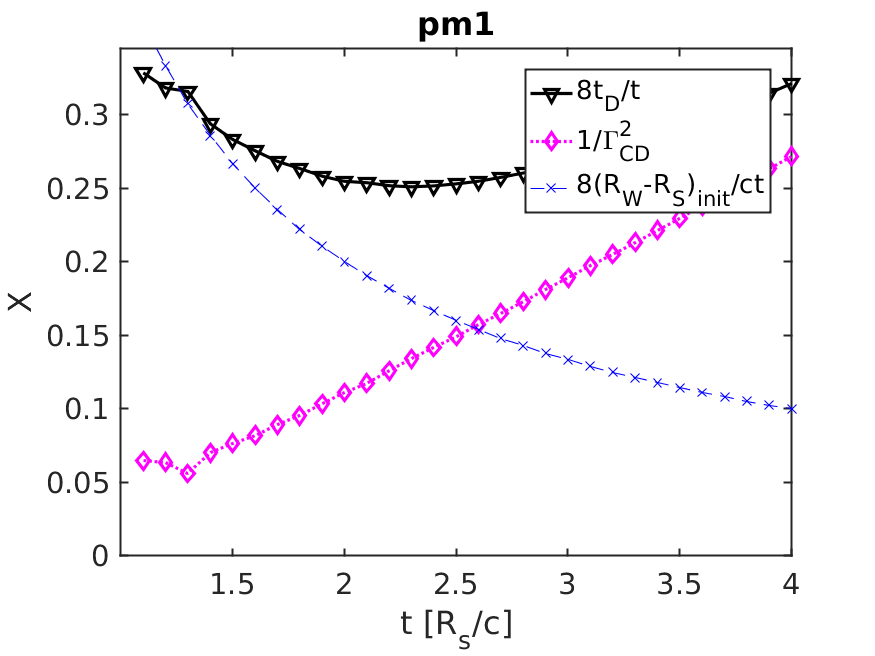}\\
\includegraphics[width=0.48\linewidth,angle=-0]{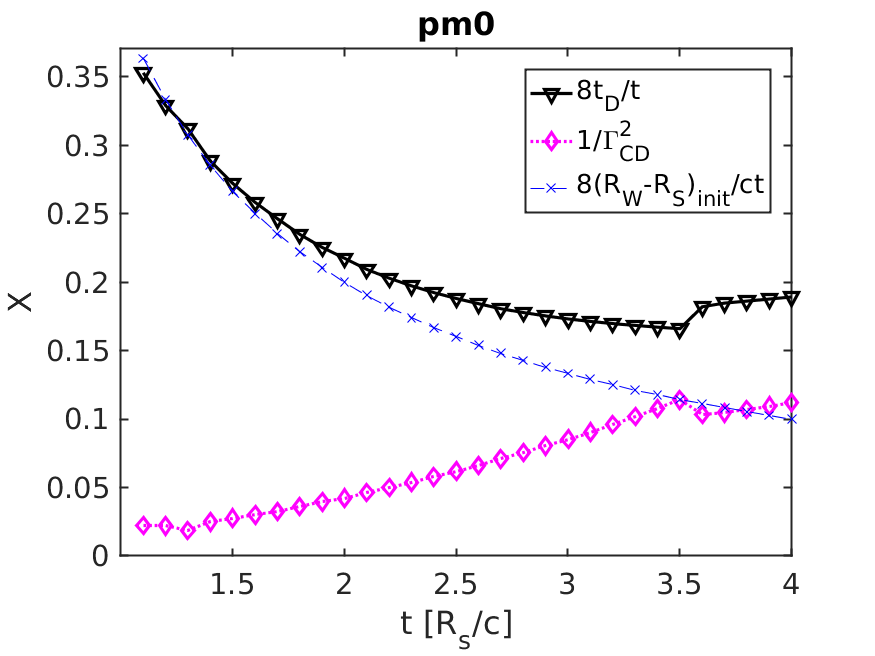}
\includegraphics[width=0.48\linewidth,angle=-0]{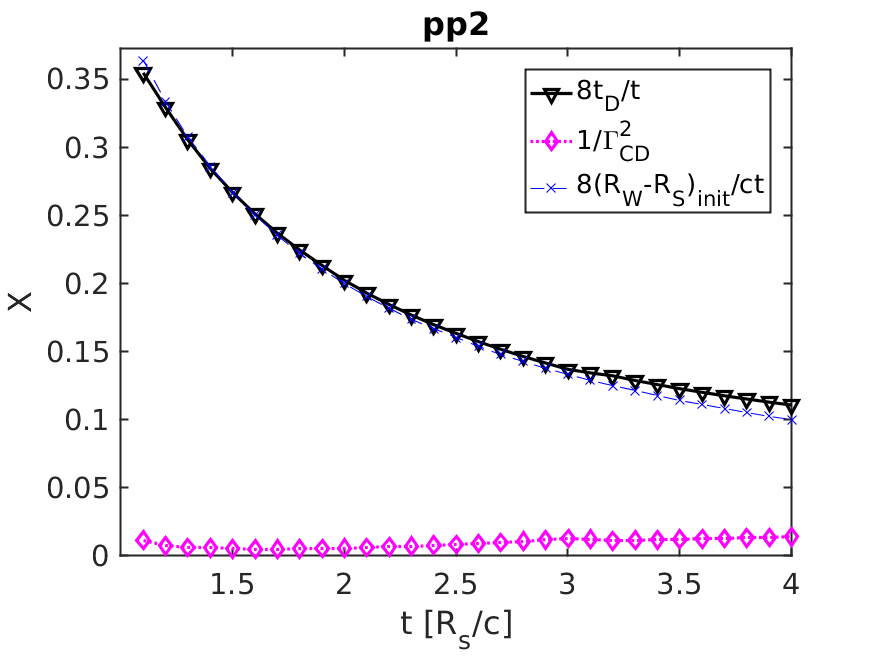}
\caption{Self-similar coordinate of the second shock $\chi$, eq~(\ref{eq:td}),  as function of time for different models. Plotted are values of  $8t_{\rm d}/t$ from simulation (triangles),
analytical curve (crosses) \protect\cite{2017PhFl...29d7101L}. Also plotted  square of  inverse Lorentz factor  (diamonds). Models with high wind power $pm0$ and $pp2$ closely follow  the theoretical curve.
}
\label{fig:tdLorCD}
\end{figure*}

\begin{figure*}
\includegraphics[width=0.48\linewidth,angle=-0]{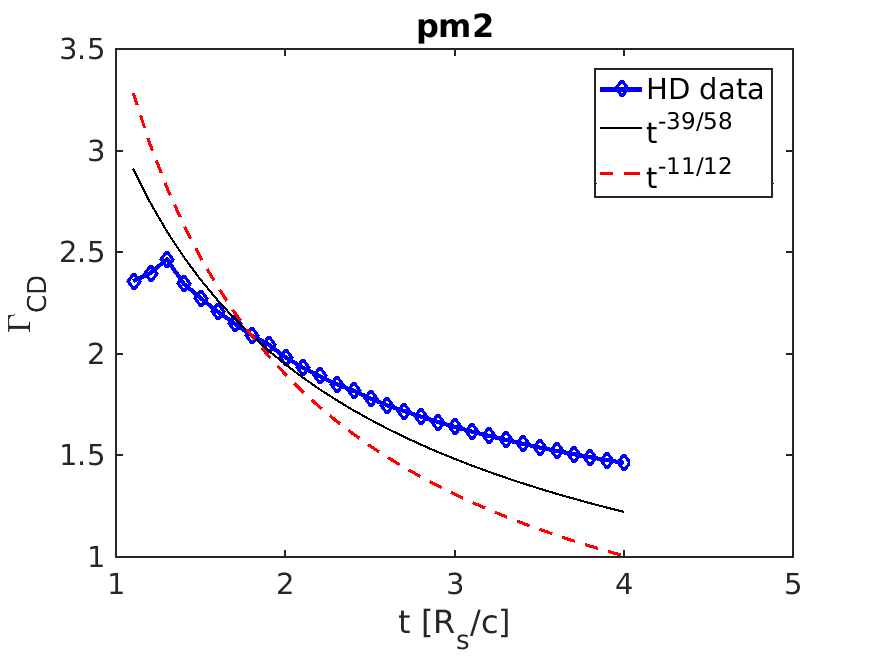}
\includegraphics[width=0.48\linewidth,angle=-0]{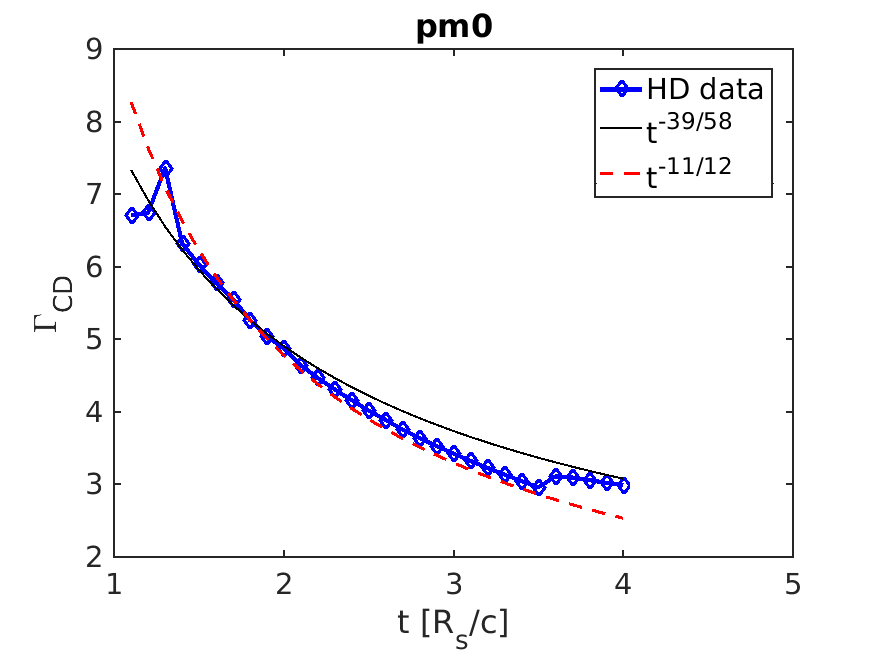}\\
\includegraphics[width=0.48\linewidth,angle=-0]{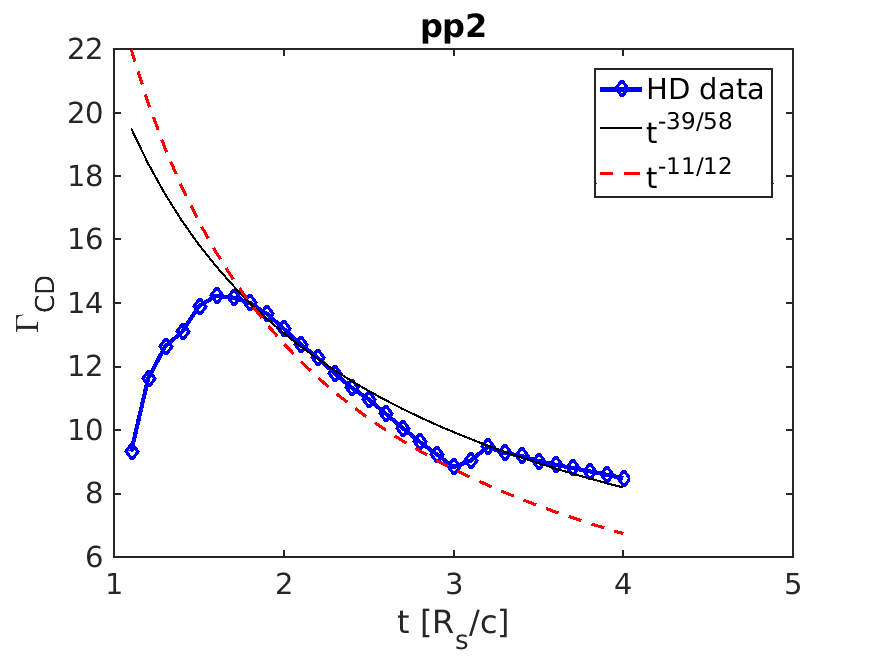}
\includegraphics[width=0.48\linewidth,angle=-0]{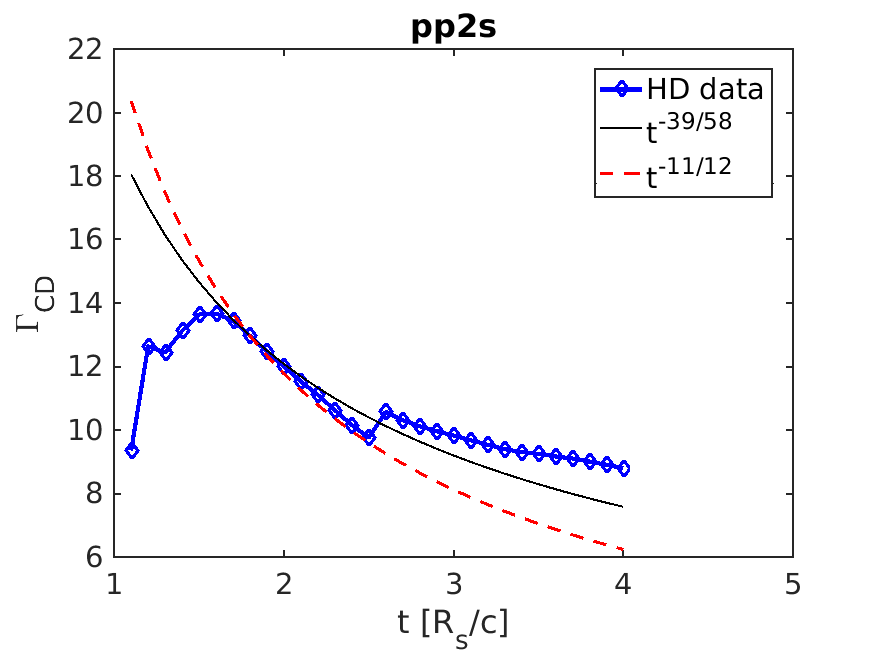}
\caption{Lorentz factor of the CD as function on time -- triangles and analytical expectations \protect\cite{2017PhFl...29d7101L}. The jumps in the \Lf\ at later times occurs  when the  wind driven FS  catches with the  leading  BMFS.
}
\label{fig:lorCD_hd}
\end{figure*}

\begin{figure*}
\includegraphics[width=0.95\linewidth,angle=-0]{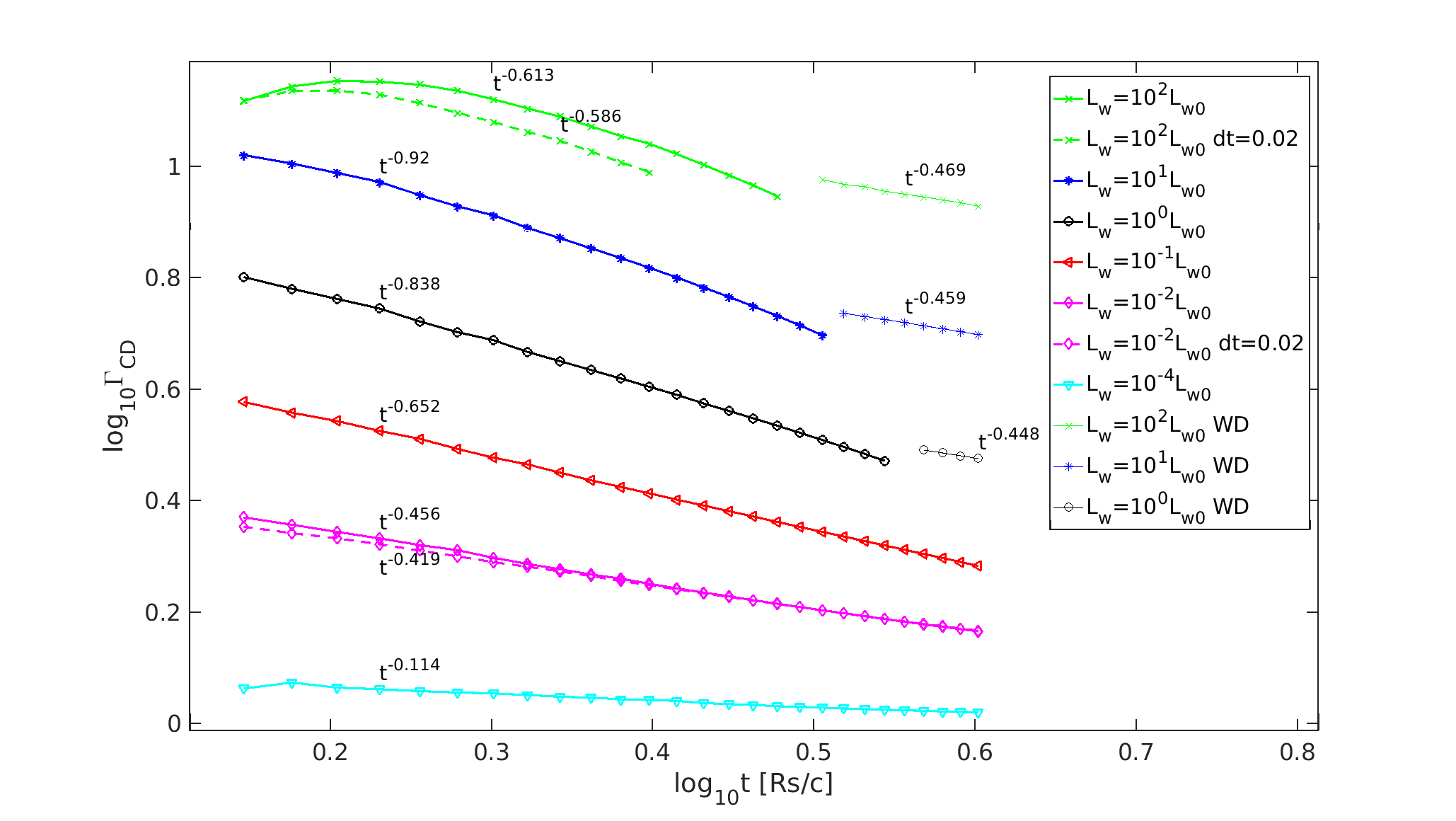}
\caption{  Lorentz factor of contact discontinuity as functions of time . The analytical estimations (see Eq.\ref{eq:1112}) $\Gamma_{\rm CD}\propto t^{-0.92}$
and for wind driven shock $\Gamma_{\rm CD}\propto t^{-0.5}$ (see thin lines with crosses, stars abd circles). { We calculate the power indexes on stright parts of the curves, $\log_{10}t>0.3 $.}}
\label{fig:LorCD_t_mhd}
\end{figure*}

\begin{figure}
\includegraphics[width=0.98\linewidth,angle=-0]{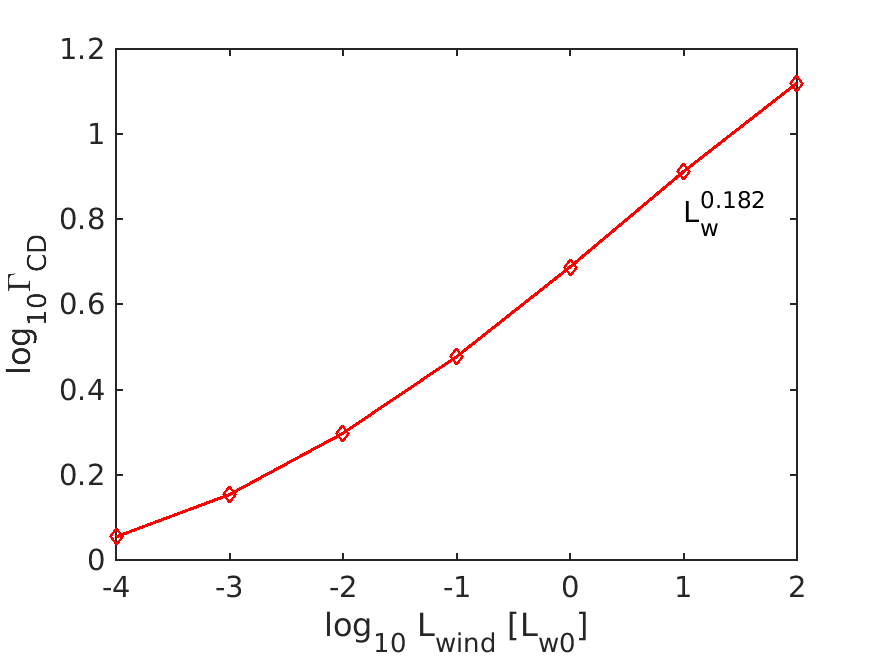}
\caption{Dependence of the Lorentz factor of the contact discontinuity  at $t=2 [R_{\rm s}/c]$. In the high wind power regime the scaling is close to the expected $\Gamma_{\rm CD} \propto L_w^{1/4}$, Eq.  (\protect\ref{eq:1112}).}
\label{fig:LorCDwp}
\end{figure}

\begin{figure*}
\includegraphics[width=0.95\linewidth,angle=-0]{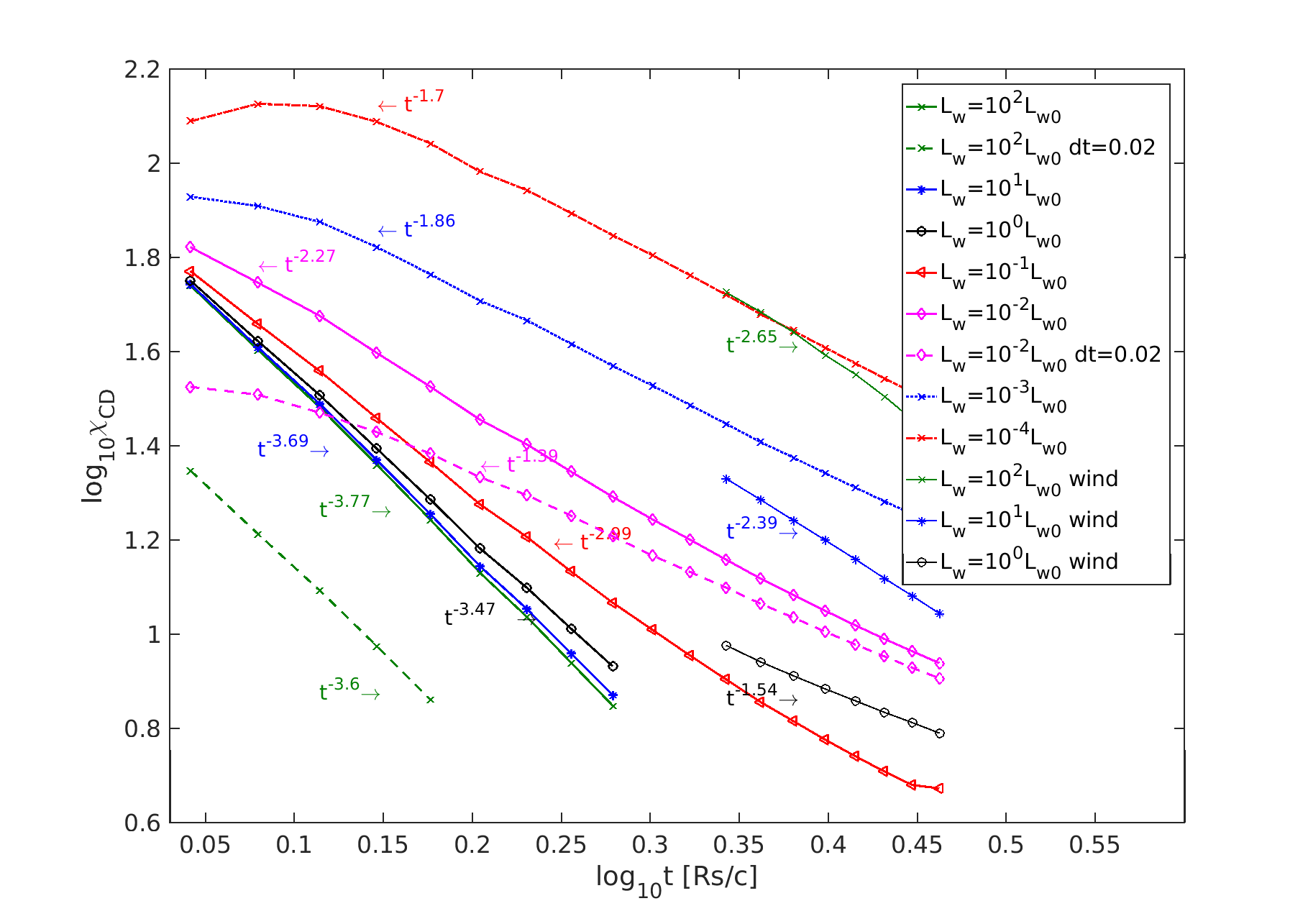}
\caption{Time dependence of  $\chi_{\rm CD}$ of contact discontinuity. { We calculate the power indexes on stright parts of the curves, $\log_{10}t>0.2$.}
} 
\label{fig:chiCD_t_mhd}
\end{figure*}

\begin{figure}
\includegraphics[width=0.98\linewidth,angle=-0]{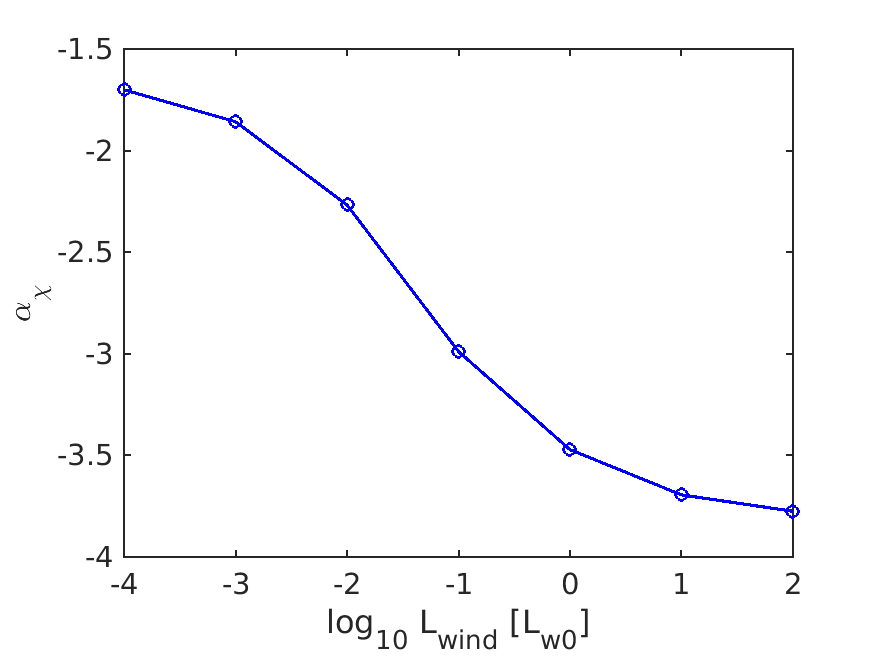}
\caption{Dependence of the   power slop  $\alpha_{\chi}$  at $t=2 [R_{\rm s}/c]$.}
\label{fig:chiCDwp}
\end{figure}

The simulations were performed using a one dimensional (1D) geometry in spherical coordinates using the {\it PLUTO} 
code\footnote{Link http://plutocode.ph.unito.it/index.html} \citep{mbm07}. 
Spatial parabolic interpolation, a 3rd order Runge-Kutta approximation in time, and an HLLD Riemann solver were used \citep{mub09}. 
{\it PLUTO} is a modular Godunov-type code entirely written in C and intended mainly for astrophysical applications and high Mach number flows in multiple spatial dimensions. The simulations were run through the MPI library in the DESY (Germany) cluster.
The flow has been approximated as an ideal, relativistic adiabatic gas with and without the toroidal magnetic field, one particle species, and polytropic
index of 4/3. The adopted resolution is $192000$ cells. The size of the domain is $r \in [0.95,4] R_{\rm s}$ or $r \in [0.98,4] R_{\rm s}$, 
here $R_{\rm s}$ is initial position of shock wave front.

As initial condition we set solution of B\&Mc with shock radius 1, Eq~(\ref{eq:bmc}), the Lorentz factor of the shock was 15. The external matter 
was assumed uniform  with density $\rho=1$ and pressure $p=10^{-4}$ (in units $c=1$). 
The pressure and density just after shock was determined 
by B\&Mc solution ($\rho_{\rm BM} = 42.43$ and $p_{\rm BM} = 150$) with total energy $E_{\rm BM} = 2.13\times 10^5$. From the left boundary (from a center)
at radius $r_{\rm w}=0.95$ or $r_{\rm w}=0.98$ (models marked by letter 's' at the end of its name) was injected wind with initial Lorentz factor $\gamma_{\rm w}=50$, the pressure of the wind was fixed $p_w = 10^{-3} \rho_w c^2$. 
The parameters of the models are listed in  Table~\ref{tab:models}.\footnote{{We change the wind density here, but power of the wind can be varied by wind Lorentz factor, magnetization or pressure. The main ingredient will be the total energy flux.}}

{The chosen setup corresponds to the following physical parameters: the density unit $n_{ISM} = 1/{\rm cm}^3$, 
total  isotropic explosion energy $E_{ISO} = 1.5\times10^{52}$~ergs, laboratory time $t_{lab} = R_s/c = 10^7$~s, the initial radius of the shock $R_s=3\times10^{17}$~cm and observer time $t_{obs} = 4.4\times10^3$~s. 
The isotropic wind power unit is $L_{w,0} = 1.2\times10^{47}$~erg/s.}

We performed nine runs without magnetic field and eight runs with different magnetizations. 
Our numerical model for the primary shock is consistent with analytical solution of BM with an accuracy  $\sim10$\% 
(pressure, density and maximal Lorentz factor).
On the top of each panel of Figures~(\ref{fig:LorTr})--(\ref{fig:tdLorCD_mhd}) we indicate name of the model with parameters presented in the Table~\ref{tab:models}.

\subsection{Results:  long-term dynamics of double explosions}
\label{s:res}

\subsubsection{Unmagnetized secondary wind}
\label{s:duw}

In the unmagnetized models labeled pXX, we vary wind density. The wind density vary from $10^{-4}$ for pm4 model to $10^2$ for pp2. 
 In Figure~(\ref{fig:LorTr})    we  plot the results of 	pXX models there we vary power of hydrodynamical wind.   At small radius one can clearly identify the  location 
of the reverse shock (RS), where the  Lorentz factor suddenly drops. At larger radius the contact discontinuity (CD) is identified by the 
the position of the tracer drop. Further out is the  secondary forward shock, and the initial BM shock. { More curves can be seen in the Appendix~\ref{sec:a1}.}

In Figure~\ref{fig:tdLorCD}   three curves are shown for pXX models: (i) theoretical curve based on the expectation from the initial conditions 
$t_{\rm d}=(r_{\rm s}-r_{\rm w})/c$; (ii) Inverse square of Lorentz factor;  (iii) actual time of delay calculated from position of
CD and its Lorentz factor using eq~(\ref{eq:td}). As we can see in the models $pm0$, $pp1$ and $pp2$ (power of the wind comparable 
to initial explosion) theoretical and actual curves are close. More powerful wind ($L_{\rm w}r_{\rm s}/c \ge 0.1E_{\rm BM}$) 
can push CD much faster that allows to satisfy conditions (\ref{eq:lorcdcon}). Large value of $\Gamma_{\rm CD}$ also relax applicability 
condition of (\ref{eq:1112}). So similar picture we can see on  Figure~\ref{fig:lorCD_hd}, here models $pp2,pp1$ and $pm0$
follow theoretically predicted time dependence (see eq~(\ref{eq:1112})) $\Gamma_{\rm CD}\propto t^{-11/12}$. 
Deviations from theoretical curves on Figures~(\ref{fig:tdLorCD}) and (\ref{fig:lorCD_hd}) at the late time are due to the fact that the wind-triggered FS reach the radius of BMFS, affecting the motion of the initial shock: in this case transition to   wind-driven 
BM solution occurs. The Lorentz factor is fitted by power law $\Gamma_{\rm CD}\propto t^{-0.45}$.

\begin{table}[]
\centering
\caption{Parameters of the models}
\label{tab:models}
\begin{tabular}{lcccc}
\tableline
\tableline
  Model              & $\rho_{\rm w}$      &   $r_{\rm w}$  &   $\sigma_{\rm w}$ &  $L_{\rm w}$ [$L_{w,0}$] \\
\tableline
&&&&\\[-5pt]
 $ pm4  $     & \quad $10^{-4}$ &  \quad 0.95 &\quad 0 & \qquad $10^{-4}$ \\
 $ pm3  $     & \quad $10^{-3}$ &  \quad 0.95 &\quad 0 & \qquad $10^{-3}$ \\
 $ pm2  $     & \quad $10^{-2}$ &  \quad 0.95 &\quad 0 & \qquad $10^{-2}$ \\
 $ pm2s $     & \quad $10^{-2}$ &  \quad 0.98 &\quad 0 & \qquad $10^{-2}$ \\
 $ pm1  $     & \quad $10^{-1}$ &  \quad 0.95 &\quad 0 & \qquad $10^{-1}$ \\
 $ pm0  $     & \quad 1         &  \quad 0.95 &\quad 0 & \qquad $1$ \\
 $ pp1  $     & \quad $10^{1}$  &  \quad 0.95 &\quad 0 & \qquad $10^{1}$ \\
 $ pp2  $     & \quad $10^{2}$  &  \quad 0.95 &\quad 0 & \qquad $10^{2}$ \\
 $ pp2s $     & \quad $10^{2}$  &  \quad 0.98 &\quad 0 & \qquad $10^{2}$ \\
\tableline
&&&&\\[-5pt]
$ mm1p1  $    & \quad $10^{1}$  & \quad 0.95 &\quad 0.1 & \qquad $11$ \\
$ m0p1  $    & \quad $10^{1}$   & \quad 0.95 &\quad 1.0 & \qquad $20$ \\
$ m05p1  $    & \quad $10^{1}$  & \quad 0.95 &\quad 3.0 & \qquad $40$ \\
$ m1p1  $    & \quad $10^{1}$   & \quad 0.95 &\quad 10  & \qquad $110$ \\
\tableline
&&&&\\[-5pt]
$ mm1ep1 $    & \quad $9.09$  & \quad 0.95 &\quad 0.1 & \qquad $10^{1}$ \\
$ m0ep1  $    & \quad $5.00$  & \quad 0.95 &\quad 1.0 & \qquad $10^{1}$ \\
$ m05ep1 $    & \quad $2.50$  & \quad 0.95 &\quad 3.0 & \qquad $10^{1}$ \\
$ m1ep1  $    & \quad $0.91$  & \quad 0.95 &\quad 10 & \qquad  $10^{1}$ \\
\tableline
&&&&\\[-5pt]
\end{tabular}
\end{table}

Figure~\ref{fig:LorCD_t_mhd}  shows time dependence of Lorentz factor at CD and its $\chi_{\rm CD}$. For high relative wind power the slope of 
Lorentz factor coincide with theoretical one. Moreover, dependence of the theoretical Lorentz factor on wind power (see eq~(\ref{eq:1112})) 
$\Gamma_{\rm CD}\propto  L_{\rm w}^{0.25}$ and simulated one (Figure~\ref{fig:LorCDwp}) 
$\Gamma_{\rm CD}\propto L_{\rm w}^{0.18}$)  are in a good agreement.

Time behavior of theoretically predicted $\chi_{\rm CD}$ ($\chi_{\rm CD}\propto t^{\alpha_\chi}$, $\alpha_\chi=-4$) is in a good agreement with models with high relative wind power, 
see Figures~(\ref{fig:chiCD_t_mhd})
and
(\ref{fig:chiCDwp})
which shows tendency of power slop to $\alpha_\chi={-3.8}$ at large wind powers.
After the moment than wind driven FS reaches BMFS, the slope is changed and tends to   $\alpha_\chi= {-2.7}$.

{The deviation from theoretically predicted slop $\Gamma_{\rm CD}\propto t^{-0.92}$ take   place when 
wind power is low. The low power wind forms sub-relativistic shock, which pushes sub-relativistic CD. Since the analytic  theory is applicable in ultra relativistic regime, this  explains   
deviations of numerical results from theory for  $\Gamma_{CD} < 3$.  The same effect works for dependence of $\chi_{\rm CD}$ on time. Sub-relativistic motion of CD can have only small values of $\alpha_\chi\sim-2$, 
and powerful winds in relativistic regime shows good agreement with theoretical predictions. }

\subsubsection{Magnetized secondary wind}
\begin{figure*}
\includegraphics[width=0.48\linewidth,angle=-0]{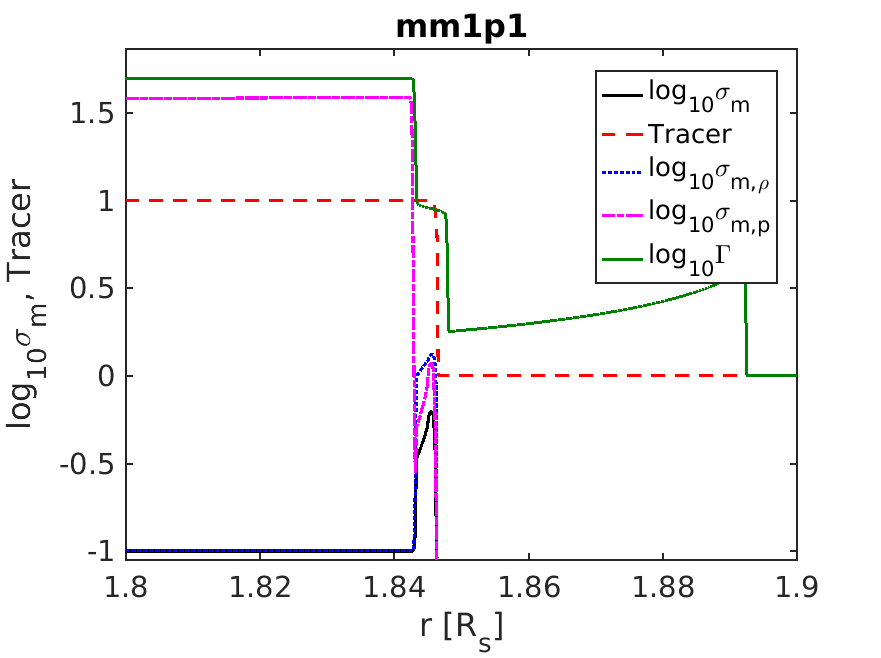}
\includegraphics[width=0.48\linewidth,angle=-0]{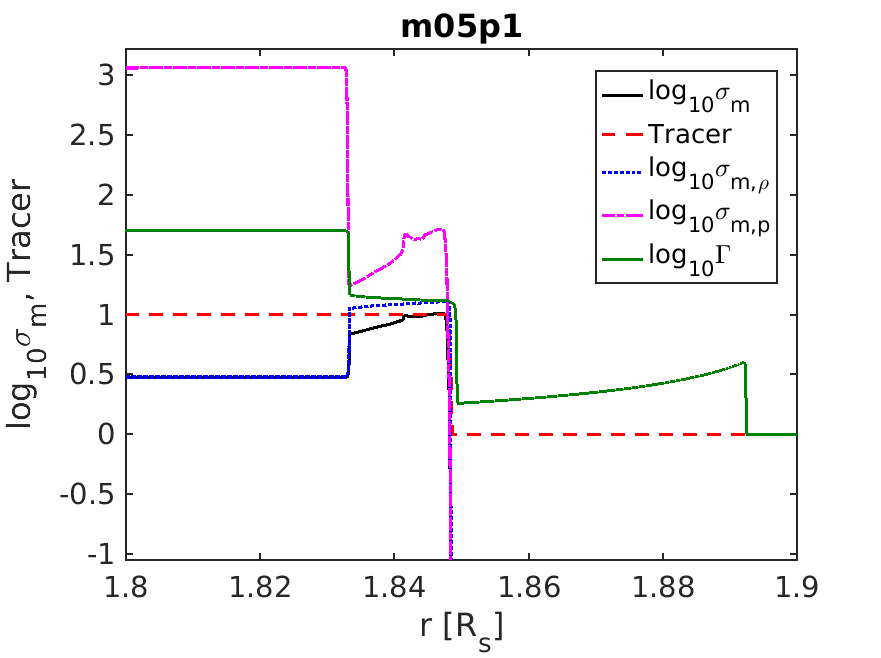}\\
\includegraphics[width=0.48\linewidth,angle=-0]{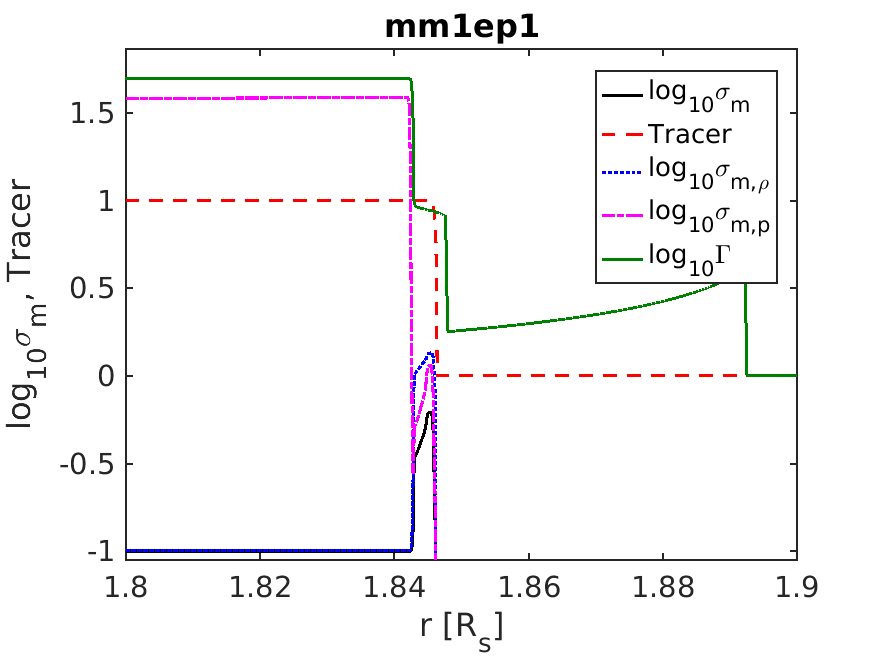}
\includegraphics[width=0.48\linewidth,angle=-0]{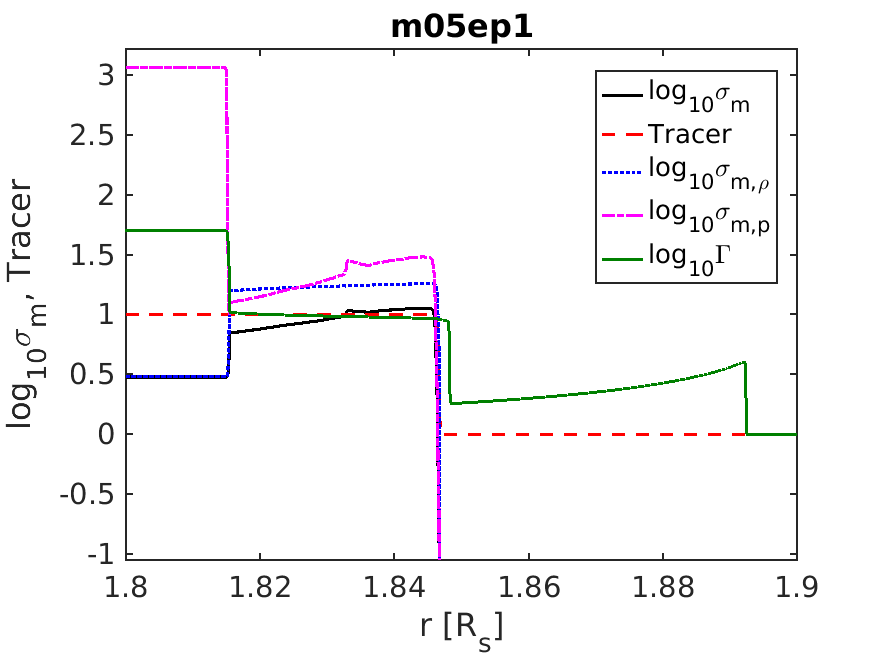}
\caption{Magnetization, tracer and Lorentz factor distribution for magnetized models. As theory predicts, the thickness of reverse shock region increase with magnetization.  }
\label{fig:sigmam_mhd}
\end{figure*}

Magnetized models marked as mXXp1 have constant wind density, where XX indicates magnetization of the flow.  Magnetized models marked as mXXep1 have constant wind luminosity, where XX 
indicates magnetization of the flow.
As a basis for the magnetized wind models, we choose  the model $pp1$, which have $L_{\rm w}r_{\rm s}/c \approx E_{\rm BM}$, so that the total wind power  injected during simulation is 
compatible to the energy of the initial explosion. 
Figure (\ref{fig:sigmam_mhd}) demonstrates the  structure of the solution. The main difference from the unmagnetized models is that  the thickness of a layer between FS and RS  increases 
with magnetization, { the similar conclusion was obtained by \cite{2009A&A...494..879M}.} This is related   to a decrease of  compressibility of the magnetized matter. Also note, that  in models with similar  total power of the wind, the position of FS 
almost independent of magnetization, while 
 the position of RS strongly depends on the wind magnetization, RS moves slower in highly magnetized models. {More solution profiles can be found in the Appendix~\ref{sec:a2}.}

\begin{figure*}
\includegraphics[width=0.48\linewidth,angle=-0]{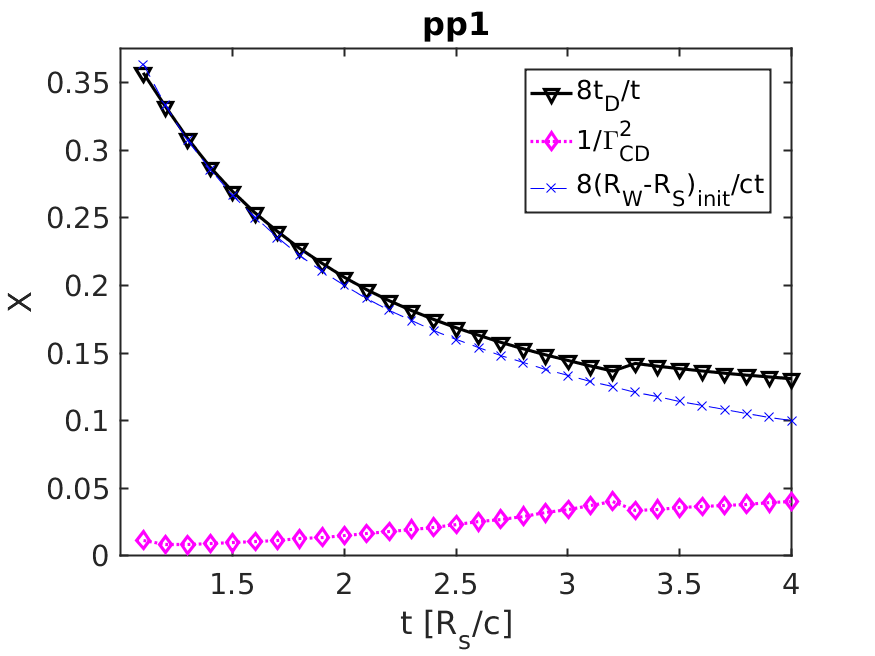}
\includegraphics[width=0.48\linewidth,angle=-0]{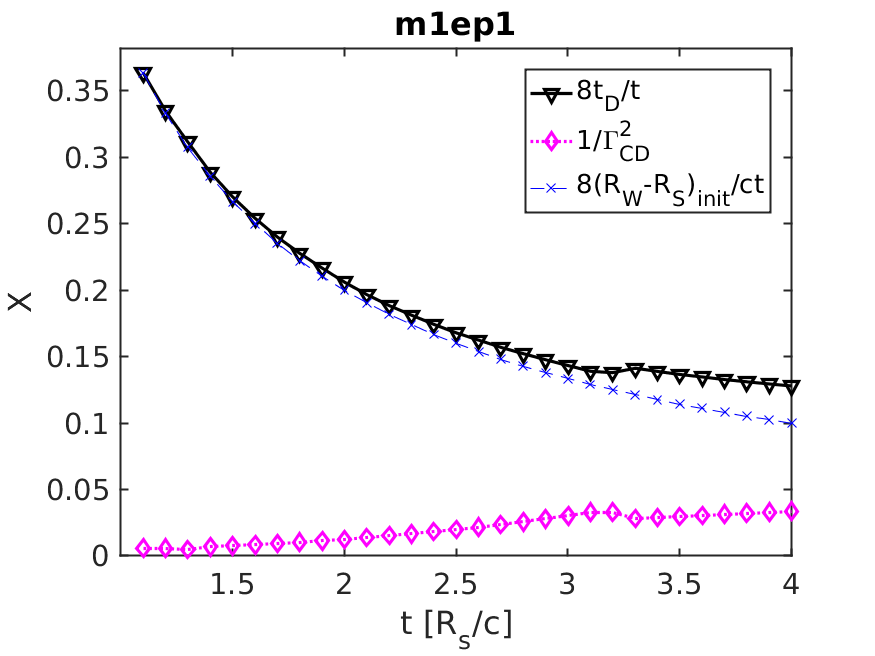}\\
\includegraphics[width=0.48\linewidth,angle=-0]{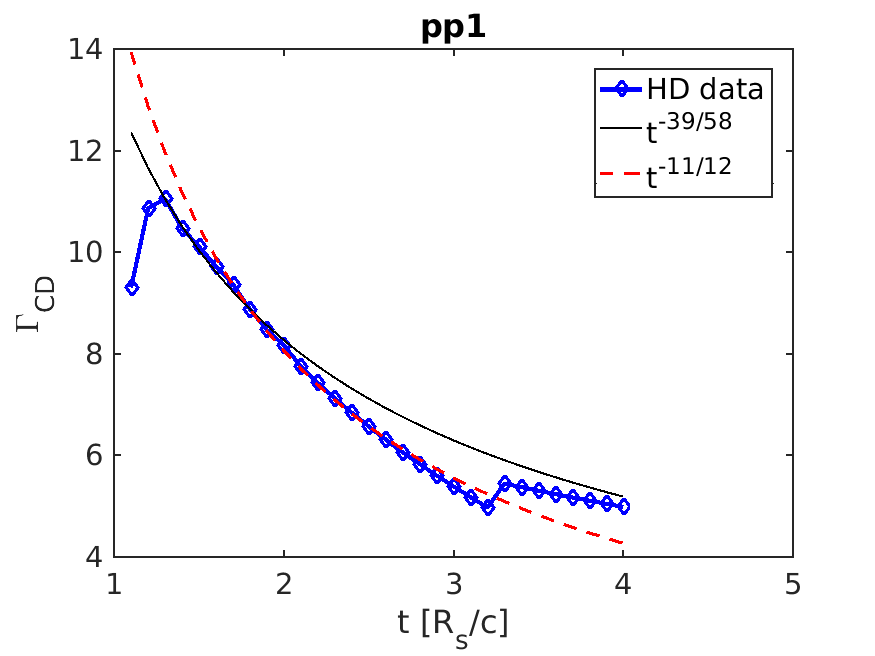}
\includegraphics[width=0.48\linewidth,angle=-0]{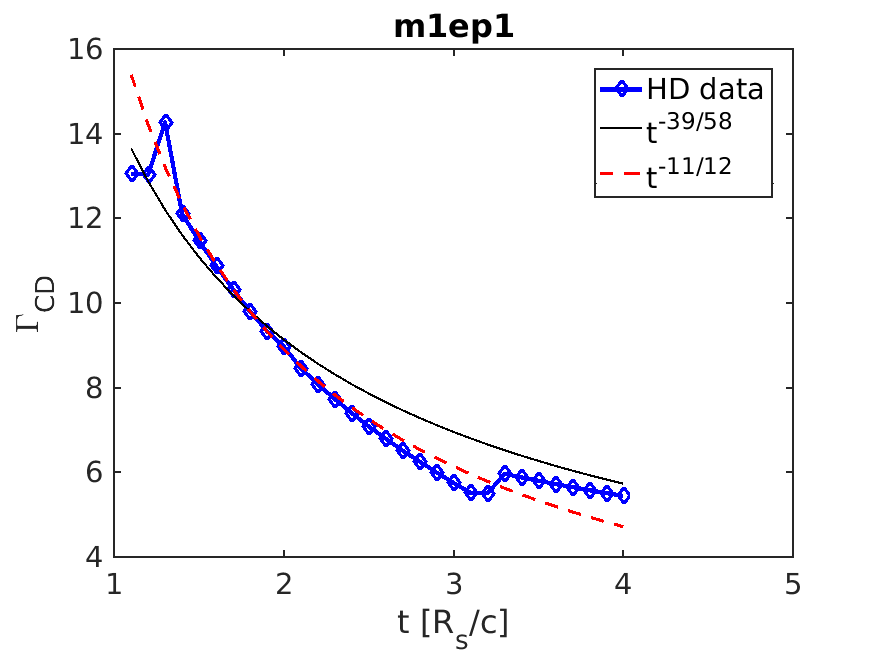}
\caption{Effects of magnetization on flow dynamics. Top row: self-similar coordinate of the second shock $\chi$ (same as Figure~\ref{fig:tdLorCD}) for cases with magnetization $\sigma_{\rm w} = 0$ (left) and $\sigma_{\rm w} = 10$ (right). 
Bottom row:  Lorentz factor as a function of  time -- triangles and analytical expectations  \protect\cite{2017PhFl...29d7101L} for cases with magnetization $\sigma_{\rm w} = 0$ (left) 
and $\sigma_{\rm w} = 10$ (right). The jumps in the \Lf\ at later times occurs  when the  wind driven FS  catches with the  leading  BMFS.
}
\label{fig:tdLorCD_mhd}
\end{figure*}

All magnetized wind models show good agreement between theoretical expectation $t_{\rm d}$ and actual ones, see Figure~\ref{fig:tdLorCD_mhd}. The Lorentz factor of CD is  also nicely fitted by theoretical curve eq.~(\ref{eq:1112}).

\begin{figure*}
\includegraphics[width=0.48\linewidth,angle=-0]{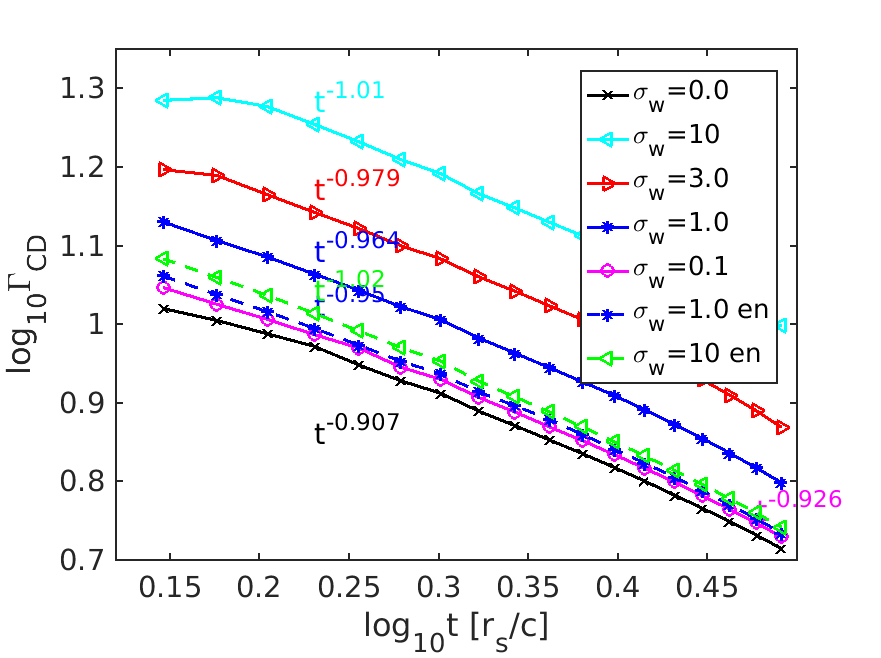}
\includegraphics[width=0.48\linewidth,angle=-0]{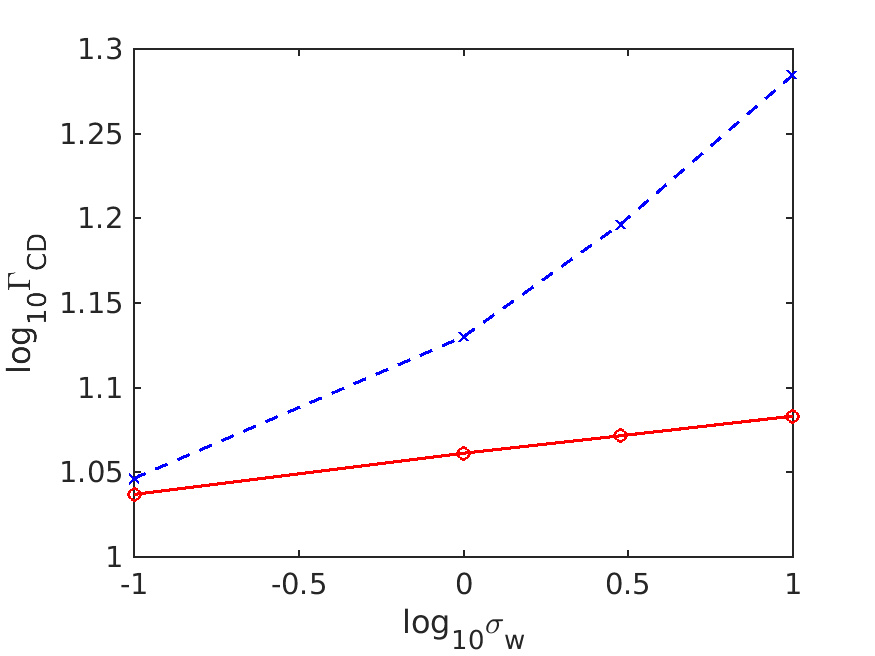}
\caption{Lorentz factor of the contact discontinuity as function of  time, left panel,  {(\cf\ Eq. (\protect\ref{eq:1112}) $\Gamma_{\rm CD}\propto t^{-0.92}$)}. 
{We calculate the power indexes on stright parts of the curves, $0.25<\log_{10}t<0.45$.}
Dependence of the  Lorentz factor of the contact discontinuity on wind magnetization at $t=1.4 [R_{\rm s}/c]$, right panel.  
Red curve is constant total power, blue dashed curve is constant matter power. As expected, in for fixed total power the \Lf\ of the CD is approximately independent of 
the the wind magnetization.
}
\label{fig:mlorCD}
\end{figure*}

The power of the slope of Lorentz factor of CD is in good agreement with theoretical one for wind independent 
on its magnetization see Figure~\ref{fig:mlorCD}. Moreover, Lorentz factor of CD very weakly depends on magnetization. If power of the wind is 
conserved $\Gamma_{\rm CD}\propto \sigma_{\rm w}^{0.023}$, if we preserve hydrodynamical part of the flow and increase magnetization trough
increasing magnetic flux, we get  $\Gamma_{\rm CD}\propto \sigma_{\rm w}^{0.18}$ that is similar to response of   $\Gamma_{\rm CD}$ on increase 
of wind power.

\begin{figure*}
\includegraphics[width=0.48\linewidth,angle=-0]{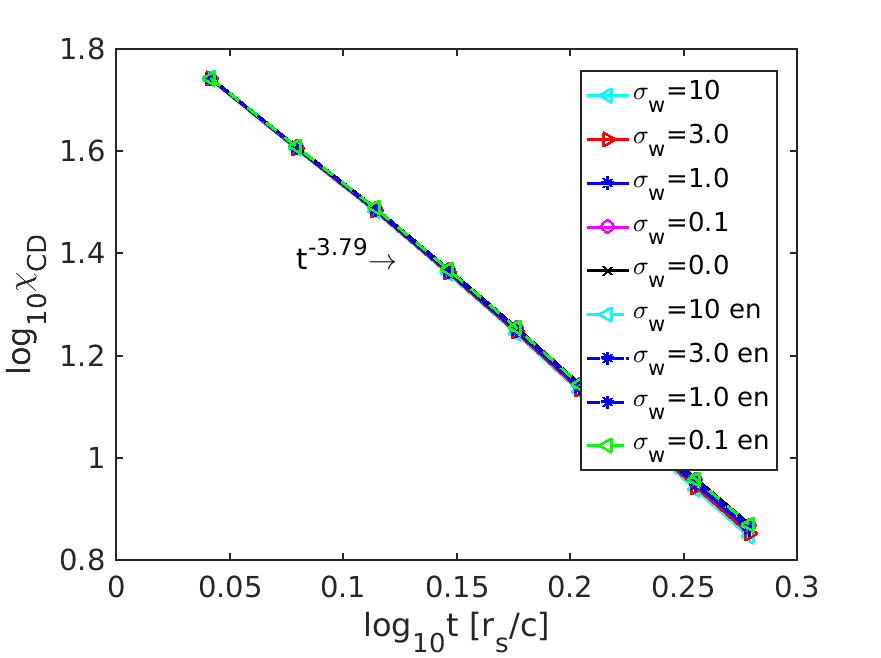}
\includegraphics[width=0.48\linewidth,angle=-0]{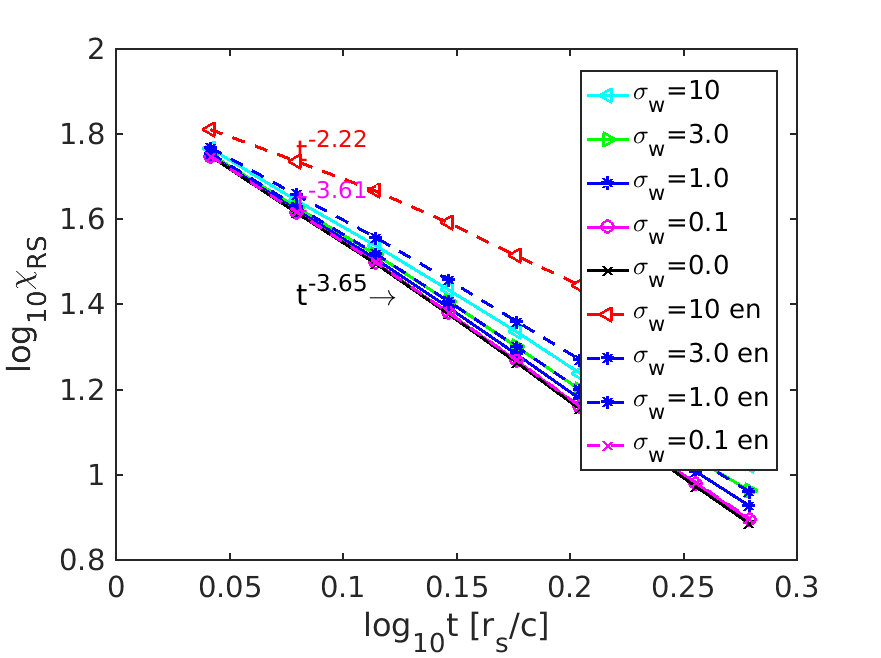}
\caption{Time dependence of  $\chi_{\rm CD}$  ({\cf} Eq.  (\protect\ref{eq:chit})) and $\chi_{\rm RS}$ (location of the CD and RS in self-similar coordinate). 
{In a fully self-similar regime the dynamics of the RS follows that of the CD.  The low $\sigma $ models do show this property. As we discussed above in the case of the CD, for smaller wind powers the effective time delay $t_d$ starts to become important, resulting in smaller  temporal indecies. We attribute flatter dependence of $\chi_{\rm CD}$ on time (see also Figure \protect \ref {fig:malchi}) to a somewhat similar effect: for larger $\sigma$ the RS \Lf\ is smaller, $ \propto \Gamma_{\rm CD}/\sqrt{\sigma}$. Thus, beyond some value of $\sigma$ the \Lf\ of the  RS and the correspoding $\chi_{\rm CD}$ are  demonstrate flatter temporal profiles.} 
}
\label{fig:mchicdrs}
\end{figure*}

\begin{figure}
\includegraphics[width=0.90\linewidth,angle=-0]{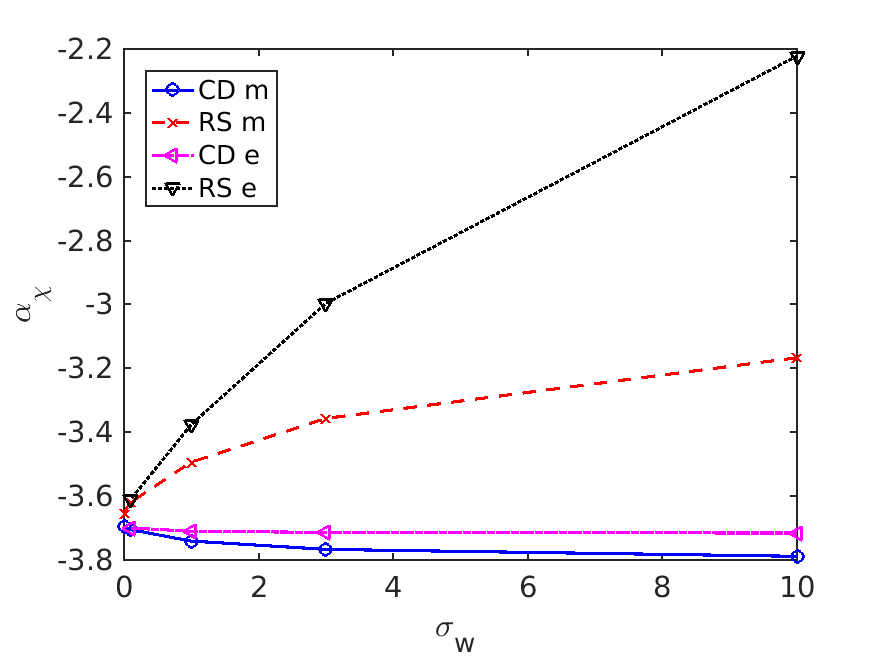}
\caption{Dependence of  $\alpha_{\chi}$ on magnetization of the wind ($\chi_{\rm CD}\propto t^{\alpha_\chi}$). 
Diamonds and crosses correspond to contact discontinuity and reverse shock in the case of a preserved energy flux of hydrodynamical flux in the wind.
Right triangle and inverted triangle correspond to the case of preserved total energy flux in the wind. (See caption for Figure \protect\ref{fig:mchicdrs}.)}
\label{fig:malchi}
\end{figure}

The power slope of time dependents of $\chi_{\rm CD}$, $\alpha_{\rm CD}$ (see  Figure~\ref{fig:mchicdrs}) almost do not depends on wind magnetization, Figure~\ref{fig:malchi}, 
and its value close to theoretically predicted slope of  $-4$.

\section{Emission from relativistic termination shock:  flares, plateaus and steep decays}
\label{rad}

Next we perform analytical calculations of expected emission properties of highly magnetized RSs. We assume that particles are accelerated at the RS, and then experience radiative and adiabatic decays. 
 In  \S \ref{Evolutionofthedistributionfunction} 
we calculate 
evolution of  the distribution function for particle injected at the shock. General relations for the observed intensity are calculated in 
 \S \ref{Observedintensity}.

\subsection{Evolution of  the distribution function}
\label{Evolutionofthedistributionfunction}

 As discussed above, the dynamics of the second shock depends on the internal  structure of the post-first shock flow, and  the wind power; all  relations are highly complicated by the relativistic and time-of-flight  effects.   To demonstrate the essential physical effects most clearly, we assume a simplified dynamics of the second shock, allowing it to propagate with constant velocity. Thus, in the frame of the shock, the \Bf\ decreases linearly with time,
\be
B^{\prime}=B_0^{\prime} \frac{t_0^{\prime}}{t^{\prime}}
\label{B}
\ee
where time $t_0^{\prime}$ and  magnetic field  $B_0^{\prime}$ are some constants.
In the following, we  assume that the RS starts to accelerate particles at time $t_0^{\prime}$, and  we calculate the emission properties of particles injected at the wind termination shock taking into account    radiative and adiabatic losses.

As the  wind generated by the long-lasting engine  starts to interact with the tail part of the flow generated by the initial explosion, the RS forms in the wind, see Figure \ref{Shock-structure-gamma}. Let's assume that the RS accelerates particles with  a power-law distribution, 
\begin{eqnarray}
f\left(\gamma^{\prime},t_i^{\prime}\right) \propto   {\gamma^{\prime}}^{-p} \Theta(\gamma^{\prime}- \gamma_{\text{min}}^{\prime})
\end{eqnarray}
where $t_i'$ is the injection time, $\Theta$ is the step-function, $\gamma'$ is the Lorentz factor of the particles, and $\gamma_{\text{min}}^{\prime}$ is the minimum Lorentz factor of the injected particles; primed quantities are measured in the flow frame. The minimal \Lf\  $\gamma_{\text{min}}^{\prime}$ can be estimated as \citep{1984ApJ...283..710K}
\be
\gamma_{\text{min}}^{\prime} \sim \gamma_{RS}  \sim  \gamma_{w}/2\Gamma_{RS}
\label{min}
\ee
\citep[We stress that in the pulsar-wind paradigm the minimal \Lf\ of accelerated particles  $\gamma_{\text{min}}^{\prime}$ scales differently from the matter-dominated  fireball case, where it is related to a fraction of baryonic energy $\epsilon_e$ carried by the wind, \eg][]{1998ApJ...497L..17S}

The accelerated particles produce synchrotron emission in the ever-decreasing \Bf, while also experiencing adiabatic losses. Synchrotron losses are given by the standard relations \citep[\eg][]{1999acfp.book.....L}. To take account of adiabatic losses we note that in a toroidally-dominated case the conservation of the first adiabatic invariant (constant magnetic flux through the cyclotron orbit) gives
\be
\partial_ {t^{\prime}} \ln \gamma^{\prime} = \frac{1}{2} \partial_ {t^{\prime}} \ln B^{\prime}
\ee
(thus, we assume that that \Bf\ is dominated by the large-scale toroidal field).

Using Eqn. (\ref{B}) for the evolution of the field,  the evolution of a particles' \Lf\ follows
\begin{eqnarray}  
&&
\frac{{d}\gamma^{\prime}}{{dt^{\prime}}}=-\frac{\tilde{C}_1 {B_0^{\prime}}^2 {\gamma^{\prime}}^2}{{t^{\prime}}^2}-\frac{\gamma^{\prime}}{2 t^{\prime}}
\nn && 
\tilde{C}_1=\frac{\sigma_T {t_0^{\prime}}^2}{6 {\pi  m _e c}}
\label{gammapde}
\end{eqnarray}
where $\sigma_T$ is the Thomson cross-section and $t_0^{\prime}$ is some reference  time.

Solving for the evolution of the particles' energy in the flow frame,
\begin{eqnarray}
\frac{1}{\gamma^{\prime}}=\frac{2 \tilde{C}_1 {B_0^{\prime}}^2}{3 t^{\prime}} \left(\left(\frac{t^{\prime}}{{t}_i^{\prime}}\right)^{3/2}-1\right)+\frac{1}{\gamma_i^{\prime}}\sqrt{\frac{t^{\prime}}{t_i^{\prime}}},
\end{eqnarray}
we can derive the evolution of a distribution function (the Green's function) \citep[\eg][]{1962SvA.....6..317K,1984ApJ...283..710K}
\begin{eqnarray} &&
G(\gamma^{\prime},t^{\prime},t_i^{\prime})=
\left\{ 
\begin{array}{cc}
{\gamma^{\prime}}^{-p} \left(\frac{t_i^{\prime}}{t^{\prime}}\right)^{\frac{p-1}{2}} \left(1-\frac{2}{3} \tilde{C}_1 {B_0^{\prime}}^2 \gamma_w^{\prime} \sqrt{t^{\prime}} \left(\frac{1}{{t_i^{\prime}}^{3/2}}-\frac{1}{{t^{\prime}}^{3/2}}\right)\right)^{p-2}, & {\gamma_{\text{low}}^{\prime}<\gamma^{\prime}<\gamma_{\text{up}}^{\prime}} \\
 0, & {else} \\
\end{array}
\right.
\nn &&
\frac{1}{\gamma_{\text{low}}^{\prime}} = \frac{2 \tilde{C}_1 {B_0^{\prime}}^2}{3 t^{\prime}} \left(\left(\frac{t^{\prime}}{t_i^{\prime}}\right)^{3/2}-1\right)+\frac{1}{\gamma_{min}^{\prime}}\sqrt{\frac{t^{\prime}}{t_i^{\prime}}}
\nn && 
\frac{1}{\gamma_{\text{up}}^{\prime}} = \frac{2 \tilde{C}_1 {B_0^{\prime}}^2}{3 t^{\prime}} \left(\left(\frac{t^{\prime}}{t_i^{\prime}}\right)^{3/2}-1\right)
\end{eqnarray}
where $\gamma_{\text{low}}'$ is a lower bound of Lorentz factor due to minimum Lorentz factor at injection and $\gamma_{\text{up}}'$ is an upper bound of Lorentz factor due to cooling.

Once we know the evolution of the distribution function injected at time $t_i'$,  we can use the Green's function to derive the total distribution function  by integrating over the injection times 
\be
{N}(\gamma^{\prime},t^{\prime}) \propto  \int _{t_i'}^{t'}  \dot{n} (t_i') G(\gamma^{\prime},t^{\prime},t_i^{\prime})dt_i^{\prime}
\ee
where $ \dot{n} (t_i') $ is  the injection rate (assumed to the constant below).

\subsection{Observed intensity}
\label{Observedintensity}

The intensity  observed at each moment depends on the intrinsic luminosity, the geometry of the flow, relativistic,  and time-of-flight effects \citep[\eg][]{1996ApJ...473..998F,2003NewA....8..495N,2004RvMP...76.1143P}. 

The intrinsic emissivity at time $t^\prime$ depends on the  distribution function $N$ and synchrotron power $P_\omega$:
\be
L^{\prime}(\omega^{\prime}, t^{\prime})=\int \int {N_A(\gamma^{\prime}, t^{\prime}) P_\omega(\omega^{\prime})} \, d\gamma^{\prime} dA^{\prime}
\ee
where $N_A$, the number of particles per unit area, is defined as $N_A = N / A = N / (2 \pi {r^{\prime}}^2 (1- \cos \theta_j))$,
$P(\omega^{\prime})$ is the power per unit frequency emitted by each electron, and $dA^{\prime}$ is the surface differential \citep[unlike][we do not have extra $\cos \theta$ in the expression for the area since we use volumetric emissivity, not emissivity from a surface]{1996ApJ...473..998F}.

We assume that the observer is located on the symmetry axis and that   the active part of the RS occupies angle $\theta_j$ to the line of sight.
The emitted power is then
\be
L^{\prime}(\omega^{\prime}, t^{\prime})= \int_{0}^{\theta_j} \int _{\gamma_{\min }^{\prime}}^{\infty }  N_A(\gamma^{\prime} ,t^{\prime}) P(\omega^{\prime}) d\gamma^{\prime}  2 \pi  {r^{\prime}}^{2} \sin(\theta) d\theta
\label{luminosity}
\ee

Photons  seen  by a distant observer at times $T_{ob}$ are emitted at  different radii and  angles $\theta$. To take account of the time of flight effects, we note that the distance between the initial explosion point and an emission point $(r^{\prime}, \theta)$ is $r^{\prime}= v t^{\prime}=v T_{ob} (1-\beta \cos(\theta))^{-1} \gamma_{RS}^{-1}$, where $T_{ob}$ is the observed time. Supposed that a photon was emitted from the distance $r^{\prime}$ and angle $\theta = 0$ at time $t^{\prime}$, and at the same time, the other photon was emitted from the distance $r^{\prime}$ and any arbitrary angle $\theta = \theta_i < \theta_j$. These two photons will be observed at time $T_0$ and $T_{\theta_i}$, then the relation between $T_0$ and $T_{\theta_i}$ is given by: 
\begin{eqnarray}
r^{\prime}=v t^{\prime}=\frac{v T_0}{(1-\beta)\gamma_{RS}}=\frac{v T_{\theta_i}}{(1-\beta \cos(\theta_i))\gamma_{RS}}
\label{surface}
\end{eqnarray}
where, the time $t^{\prime}$ measured in the fluid frame, and the corresponding observe time $T_{ob}$, is a function of $\theta$ and $t^{\prime}$: 
\begin{eqnarray}
T_{ob} = t \left(1 - \beta \cos \theta\right)= t^{\prime} \left(1 - \beta \cos \theta\right) \gamma_{RS}
\label{ob}
\end{eqnarray}

Taking the derivative of Eqn. (\ref{ob})  we find 
\begin{eqnarray}
\sin(\theta) d\theta = - \frac{T_{\text{ob}}}{{t^{\prime}}^2 \beta \gamma_{RS}} dt^{\prime} \approx - \frac{T_{\text{ob}}}{{t^{\prime}}^2 \gamma_{RS}} dt^{\prime}
\label{theta}
\end{eqnarray}
Substitute the relation (\ref{theta}) into  (\ref{luminosity}), the observed  luminosity becomes
\begin{eqnarray}
L^{\prime}(T_{ob},\omega^{\prime}) \approx \int_{t_{\theta^{\prime}=0}^{\prime}}^{t_{\theta^{\prime}=\theta_j}^{\prime}} \int _{\gamma_{\min }^{\prime}}^{\infty } \frac{ 2 \pi c^2 T_{ob}}{\gamma_{RS}} \times N_A(\gamma^{\prime} ,t^{\prime}) P(\omega^{\prime}) d\gamma^{\prime} dt^{\prime}
\label{drop}
\end{eqnarray}

To understand the Eqn. (\ref{drop}), the radiation observed at $T_{ob}$ corresponds to the emission angle from $0$ to $\theta_{j}$, which also corresponds to the emission time $t_{\theta^{\prime}=0}^{\prime} ={T_{ob}}/{(1-\beta)\gamma_{RS}}$ to $t_{\theta^{\prime}=\theta_j}^{\prime} ={T_{ob}}/{(1-\beta \cos \theta_j)\gamma_{RS}}$. So we need to integrate the emissivity function over the range of the emission angle, or integrate the emissivity function over the range of the emission time from $t_{\theta^{\prime}=0}^{\prime}= {T_{ob}}/{(1-\beta)\gamma_{RS}}$ to $t_{\theta^{\prime}=\theta_j}^{\prime}={T_{ob}}/{(1-\beta \cos \theta_j)\gamma_{RS}}$.

Finally, taking into account Doppler effects  (Doppler shift $\omega = \delta \omega^{\prime}$ and the intensity boost $I_{\omega} \left(\omega\right) = \delta^3 I_{\omega^{\prime}}^{\prime} \left(\omega^{\prime}\right)$; where  $\delta$ is the Doppler factor $\delta = 1/({\gamma_{RS} \left(1 - \beta \cos \theta\right)}) $), 
substitute the relation $t^{\prime}={T_{\text{ob}}}/{(1-\beta \cos(\theta))\gamma_{RS}}$ into Eqn.(\ref{drop}) we finally arrive at the equation for the observed  spectral luminosity:
\begin{eqnarray}
F_\om  = \int_{\frac{T_{\text{ob}}}{(1-\beta \cos(\theta_j))\gamma_{RS}}}^{\frac{T_{\text{ob}}}{(1-\beta)\gamma_{RS}}} \int _{\gamma_{\min }^{\prime}}^{\infty } \frac{1}{2 \gamma_{RS}}  c^2  D^{-2} T_{\text{ob}} \delta^3 N_A P(\omega/\delta) d\gamma^{\prime} dt^{\prime}
\label{final}
\end{eqnarray}
where $D$ is the distance to the GRB.

Next we apply these  general relations   to   three specific problem: (i) origin of  plateaus in afterglow light curves; (ii) sudden drops in the afterglow light curves  \S \ref{Plateaus};  (iii)  afterglow flares, \S \ref{flares1}.  For numerical estimates,  we assume the redshift $z = 1$, the Lorentz factor of the wind $\gamma_w = 5 \times 10^5$, the wind luminosity $L_w = 10^{46}$ erg/s, the initial injection time $t_0^{\prime} = 10^5$s (in jet frame), the power law index of particle distribution $p=2.2$,  and the viewing angle is 0 (observer on the axis)  for all calculations.  

\subsection{Results: plateaus and  sudden intensity drops in afterglow light curves}
\label{Plateaus}

\begin{figure}[h!]
  \centering
\includegraphics[width=0.49\textwidth]{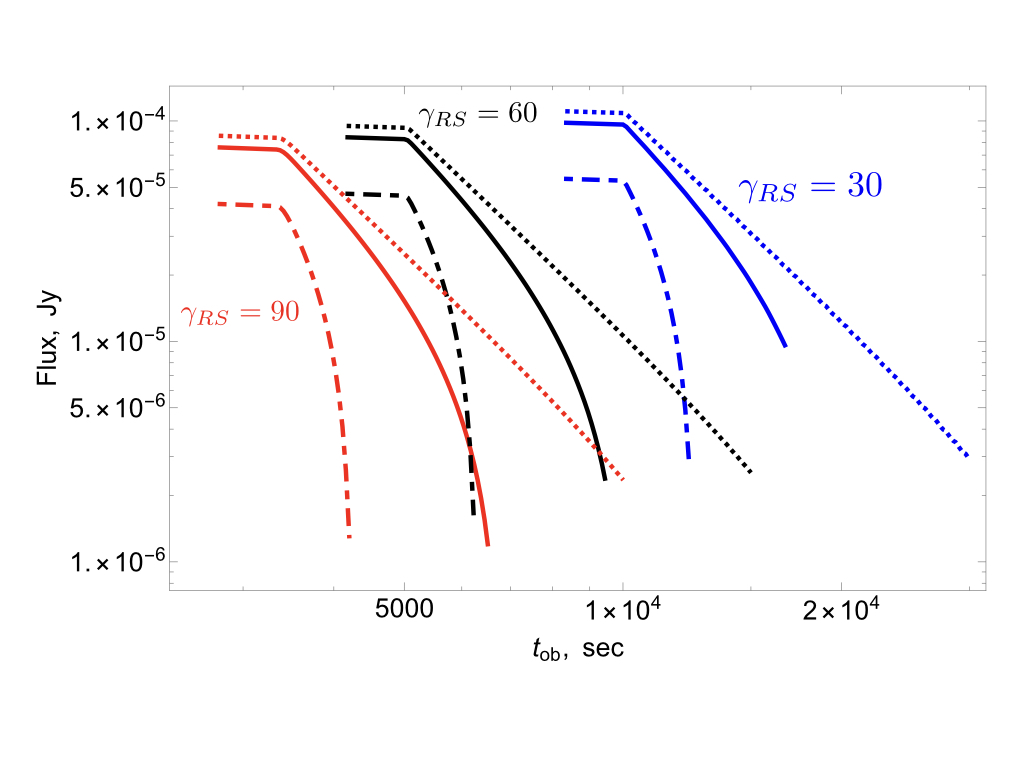}
  \caption{The light curve at 100 KeV  for different  Lorentz factors of the post-RS flow and different  jet angles $2/\gamma_{RS}$ (dotted)  $ 1/\gamma_{RS}$ (solid)  and $1/(2\gamma_{RS})$  (dot dashed) and different {\Lf}s of the RS. For $\theta_ j  \leq 1/\gamma_{RS} $  the drop in intensity is extremely fast. 
  }
  \label{GammaCD}
 \end{figure}

 Particles accelerated at the RS emit   in the fast cooling regime. The resulting synchrotron luminosity $L_s$ is approximately proportional to the wind luminosity $L_w$, as discussed by \cite{2017ApJ...835..206L}. (For highly magnetized winds with $\sigma\gg 1$ the RS emissivity is only mildly suppressed, by  high magnetization, $\propto 1/\sqrt{\sigma}$, due to the fact that higher sigma shocks propagate faster with respect to the wind.) Thus, the constant wind will produce a nearly constant light curve:  plateaus are natural consequences in our model in the case of constant long-lasting wind, see Figure \ref{GammaCD}. 
   At the early times  all light curves show a nearly constant evolution with time, a plateau,  with flux $\propto t_{ob}^{-0.1}$. A slight temporal decrease is due to the fact that \Bf\ at the RS decreases with time so that particles emit less efficiently.  This observed temporal decrease is flatter than  what is typically observed, $\propto t_{ob}^{-\alpha_2}$ with $\alpha_2=0.5-1$  \citep{2006ApJ...642..389N}. A steeper decrease can be easily accommodated due to  the decreasing wind power. This explains the plateaus.

Next we assume that  the central engine suddenly  stops operating. This process could be due to the collapse of a \NS\ into a black hole or sudden depletion of an accretion disk. At a later time, when the ``tail'' of the wind reaches the termination shock, acceleration stops. Let the 
 injection terminate  at a some  time $t_{\text{stop}}^{\prime}$. The distribution function in the shocked part of the wind then become
\begin{eqnarray}
{N}(\gamma^{\prime},t^{\prime}) \propto \int _{t_0^{\prime}}^{\min(t^{\prime},t_{\text{stop}}^{\prime})}G(\gamma^{\prime},t^{\prime},t_i^{\prime})dt_i^{\prime}
\end{eqnarray}

\begin{figure}[h!]
  \centering
  \includegraphics[width=0.49\textwidth]{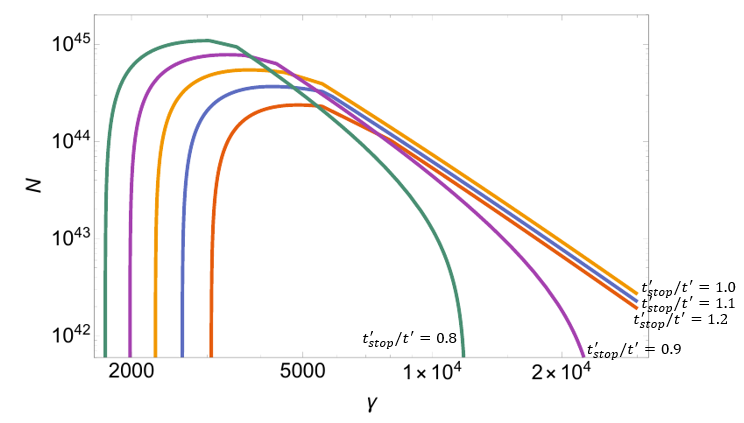}
  \caption{Evolution of the distribution function. Here we take  account the effect of radiation loss and adiabatic expansion. In our calculation, the Lorentz factor of RS $\gamma_{RS}=90$, and the injection is stopped at time $t_{\text{stop}}^{\prime} = 1.5 \times 10^5$s, $\gamma_{\min} = \gamma_w/\gamma_{RS} = 5.5 \times 10^3$, initial magnetic field $B_0 = 2.1$G. The times are measured in fluid frame at $t_{\text{stop}}^{\prime}/t^{\prime} = 1.2, 1.1, 1.0, 0.9, 0.8$ from red to green curves.}
  \label{stop_at_150000}
 \end{figure}

 Figure \ref{stop_at_150000} shows the evolution of the distribution function by  assuming the Lorentz factor of RS $\gamma_{RS}=90$, and the injection is stopped at time $t_{\text{stop}}^{\prime} = 1.5 \times 10^5$s (in this case, the $T_{\text{ob,stop}} = 833$s in the observer's frame).  The number of high energy particles drops sharply right after  the injection is stopped:   particles lose their energy via synchrotron radiation and adiabatic expansion in fast cooling regime.

The resulting light curves   are plotted in Figure \ref{GammaCD}. We assume post-RS flow $\gamma_{RS}=30, \, 60, \, 90$ and three jet opening angles of $\sim  (1/2,\, 1,\,2)  \times  \gamma_{RS}^{-1}$. These particular  choices of $\theta_j$ are  motivated by our expectation that sudden switch-off of the acceleration at the RS will lead to fast decays in the observed flux (in the fast cooling regime).

 The injection is stopped at a fixed time in the fluid frame,  corresponding to $t_0'=6 \times 10^5$s. 
   There is a sudden drop of intensity when the injection is stopped   ($T_{\text{ob}} = 10000$s for blue curve, $T_{\text{ob}} = 5000$s for black curve, and $T_{\text{ob}} = 3 \times 10^3$s for red curve).  Blue curve has $\gamma_{RS} = 30$, $\gamma_{\min} = \gamma_{\text{w}}/\gamma_{RS}= 1.6 5. \times 10^4$, initial magnetic field $B_0$ = 6.4G; green curve has $\gamma_{RS} = 60$, $\gamma_{\min} = \gamma_w/\gamma_{RS} =8.3 \times 10^3$, initial magnetic field $B_0$=3.2G; red curve has $\gamma_{RS} = 90$, $\gamma_{\min} = \gamma_w/\gamma_{RS}= 5.5 \times 10^3$, initial magnetic field $B_0$ = 2.1G. Here we assume $B_0 \propto {\gamma_{RS}}^{-1}$ for our calculations. Smaller jet angle produce sharper drop.  

In the simplest qualitative explanation, consider a shell  of radius $r_{em}$ extending to a finite angle $\theta_j$ and producing an instantaneous flash of emission (instantaneous is an approximation to the fast cooling regime). The observed light curve is then \cite{1996ApJ...473..998F}
\be 
\propto 
\left\{
\begin{array}{cc}
\left( \frac{T_{ob}}{T_0} \right)  ^ {-(\alpha +2)}, & 0< T_{ob} <  \frac{r_{em} /c} {2 } \theta_j^2 \\
0 &   \frac{r_{em} /c} {2 } \theta_j^2  <  T_{ob}
\end {array}
\right.
\ee
where $T_0 =  \frac{r_{em} /c} {2 \gamma_{RS}^2}$ and $\alpha$ is the spectral index. Thus,  for $\theta_j> 1/\gamma_{RS}$ the observed duration of a pulse is $\sim T_0$, while  for $\theta_j<  1/\gamma_{RS}$ the pulse lasts  much shorter, $\sim T_0 ( \theta_j \gamma_{RS})^2 \ll T_0$. Thus,  in this case 
a drop in intensity is faster than what would be expected in either faster  shocks or shocks producing emission in slowly cooling regime.

\subsection{Results:   afterglow  flares}
\label{flares1}

\begin{figure}[h!]
  \centering
  \includegraphics[width=0.49\textwidth]{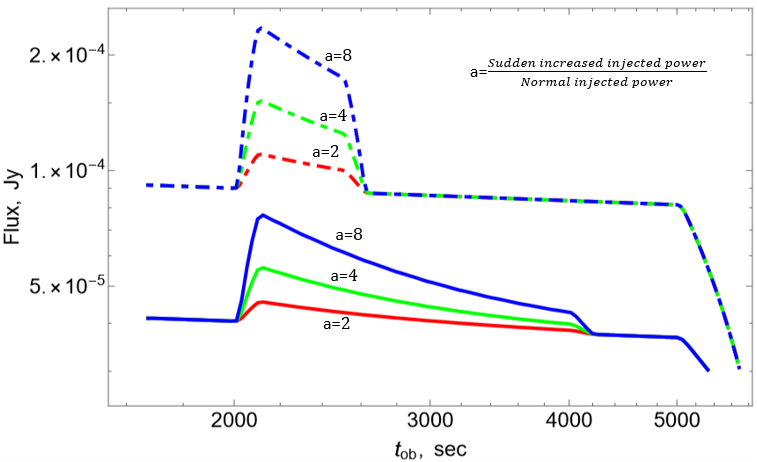}
  \caption{Afterglow flares due to variations in wind luminosity for the case  $\gamma_{RS} =60$ (green curve in the Figure \protect\ref{GammaCD}). The ejected power is increased by factors  $a=2, 4, 8$ for a short period from $2.4 \times 10^5$s to $2.5 \times 10^5$s (in the fluid frame). Solid lines are for  $\theta_j = \gamma_{RS}$, dashed lines are for  $\theta_j = 1/2\gamma_{RS}$.
 For clarity, the relative shift of intensities between the plots for two opening angles is due to our parametrization of the injected power (constant total power not isotropic equivalent).
}
  \label{flare}
 \end{figure}

Next, we investigate the possibility that  afterglow  flares are produced due to the variations in wind power. 
We re-consider the case of $\gamma_{RS} = 60$ (the green curve in Figure \ref{GammaCD}),
  but set the ejected power at two, four, and eight times larger than the  average   power for a short period of time from $2.4 \times 10^5$s to $2.5 \times 10^5$s. We consider the two cases: the wide jet angle ($\theta_j = 1/\gamma_{RS}$) and the narrow jet angle ($\theta_j = 1/2\gamma_{RS}$). The corresponding light curves are plotted in Figure \ref{flare}.

Light curves show a  sharp rise around $T_{ob} = 2000$ corresponding  to the increased ejected power $t = 2.4 \times 10^5$s at emission angle $\theta = 0$, followed by a   sharp drop around $T_{ob} = 4000$s for the case of wide jet and $T_{ob} = 2500$s for the case of narrow jet (which corresponds to the ending time of the increased ejected power $t = 2.5 \times 10^5$s at emission angle $\theta = \theta_j$). Bright flares are clearly seen. Importantly,   the corresponding total injected energy  is only  $\sim 1\%, \, 5\%$ and  $10\%$  larger than the averaged value.
 The magnitude of the rise in flux is less than the magnitude of the  rise in ejected power (e.g. the rise in ejected power by a factor eight only gives the rise in flux by a factor two), due to the fact that   the emission from the increased ejected power from different angles is spread out in observer time.
Thus, variations in the wind power, with  minor total energy input, can produce bright afterglow flares. \citep[][]{Lyutikov:2006a}

\section{Discussion}
\label{s:disc}

In this paper we discuss  properties of GRB afterglows   within the ``pulsar wind'' paradigm:   long-lasting, ultra-relativistic, highly magnetized wind with particles  accelerated at the wind termination shock \citep{1984ApJ...283..710K}. 
The present model of long lasting winds in GRBs is qualitatively different from previous models based on  ``fireball'' paradigm,  see  \S \ref{modelss}.

We first performed  a set of  detailed   RMHD simulations of   relativistic double explosions. Our numerical results are in excellent  agreement with theoretical prediction \citep{2017PhFl...29d7101L,2017ApJ...835..206L}. For example,  for sufficiently high wind power we have $\Gamma_{\rm CD}\propto t^{-11/12}$, 
while after $t_{eq}$ the shocks merge and move as a single self-similar shock with  $\Gamma_{\rm CD} \propto t^{-1/2}$. 
In addition numerics demonstrates a  much richer set of phenomena (\eg, transitions between various analytical limits and variations in the temporal slopes).  We find that even for the case of constant external density and constant wind power the dynamics of the wind termination shock shows a large variety - both 
in  temporal slopes of the scaling of the Lorentz factor of the shock, and producing  non-monotonic behavior. Non-self-similar evolution of the wind termination shock occurs for two different reasons: (i) at early times due to a delay in the activation of the long-lasting fast wind; 
(ii) at late times when the energy  injected by the wind becomes comparable to the energy of the initial explosion.

Second, we performed radiative calculations of the RS emission and we demonstrated that  emission from the long-lasting relativistic wind can resolve a number of contradicting GRB observations. We can reproduce:
\begin{itemize} 
\item  Afterglow plateaus: in the  fast cooling regime the emitted power is comparable to the wind power. Hence, only mild wind luminosity $L_{w}\sim 10^{46}$ erg s$^{-1}$ is required (isotropic equivalent)
\item Sudden drops in afterglow light  curves: if the central engine stops operating, and if at the corresponding moment the \Lf\ of the RS is of the order of the jet angle, a sudden drop in intensity will be observed.
\item Afterglow flares: if the wind intensity varies, this leads to the sharp variations of afterglow luminosities. Importantly, a total injected {\it energy} is small compared to the total energy of the explosion.
\end {itemize} 

 \cite{2017ApJ...835..206L} also discussed how the model provides explanations for a number of other GRB phenomena,  like ``Naked GRBs problem'' \citep[][]{2006ApJ...637L..13P,2008AIPC.1000..191V}  (if the explosion does not produce a long-lasting wind, then there will be no X-ray afterglow since RS reflects the properties of wind),
``Missing orphan afterglows'':
both prompt emission and afterglow emission arise from the engine-powered flow, so they may  have similar collimation properties. The model also offers explanations to missing and/or chromatic jet breaks,  orphan afterglows, ``Missing'' reverse shocks  (they are not missing - they are dominant).

In conclusion,   the high energy  emission from highly relativistic wind is (i) highly efficient; 
(ii) can be  smooth (over a period of time)  for constant wind parameters; (iii) can react quickly to the changes of the wind properties. 
RS also contributes to the  optical - this explains correlated X-optical features often seen in afterglows. FS emission occurs in the optical range, and, at later times, in radio  \citep{2017ApJ...835..206L}.

\section*{Acknowledgments}
 

We thank the {\it PLUTO} team for the possibility to use the {\it PLUTO} code and for technical support. 
The visualization of the results performed in the VisIt package \citep{HPV:VisIt}. 
This work had been supported by 
NASA grants 80NSSC17K0757 and 80NSSC20K0910,   NSF grants 10001562 and 10001521, and  NASA Swift grant 1619001

The data that support the findings of this study are available from the corresponding author upon reasonable request.

\appendix

\section{Simulation profiles}
\subsection{Non magnetized cases}
\label{sec:a1}

\begin{figure*}
\includegraphics[width=0.48\linewidth,angle=-0]{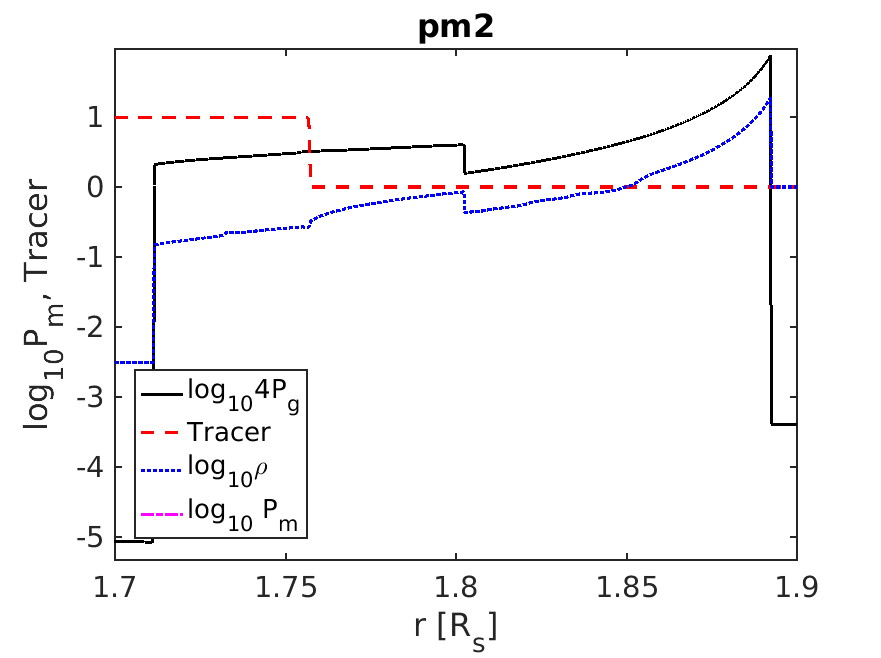}
\includegraphics[width=0.48\linewidth,angle=-0]{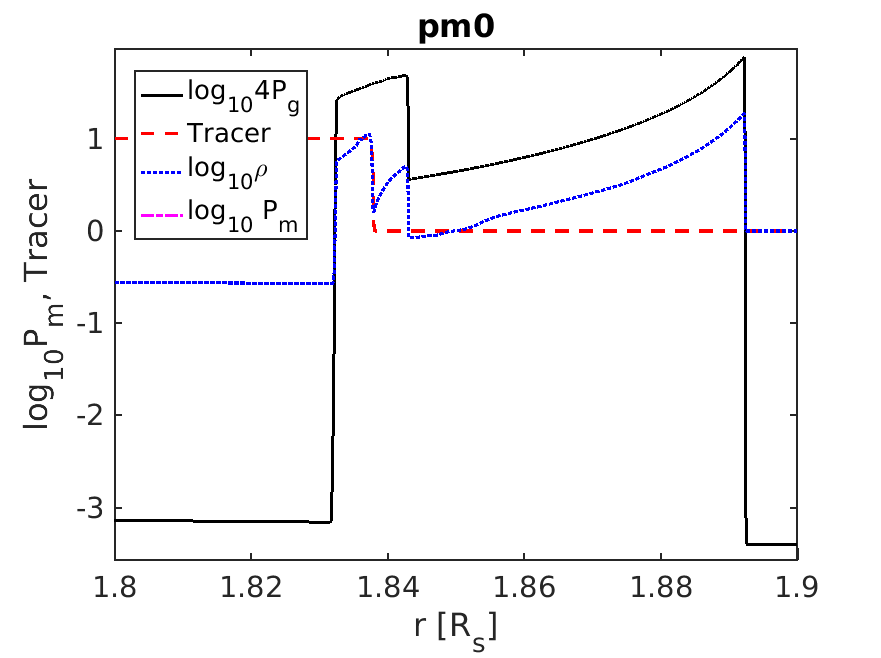}\\
\includegraphics[width=0.48\linewidth,angle=-0]{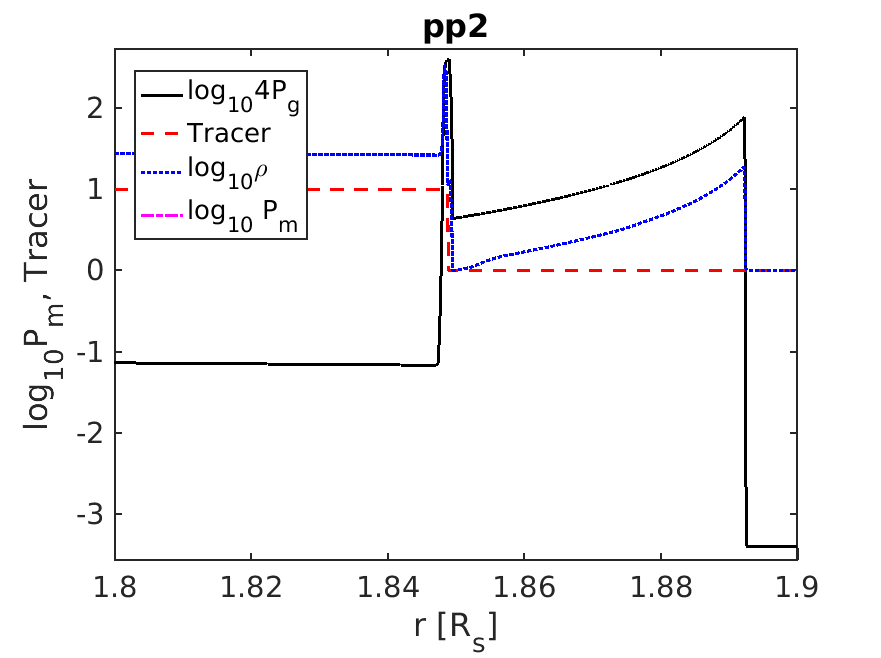}
\includegraphics[width=0.48\linewidth,angle=-0]{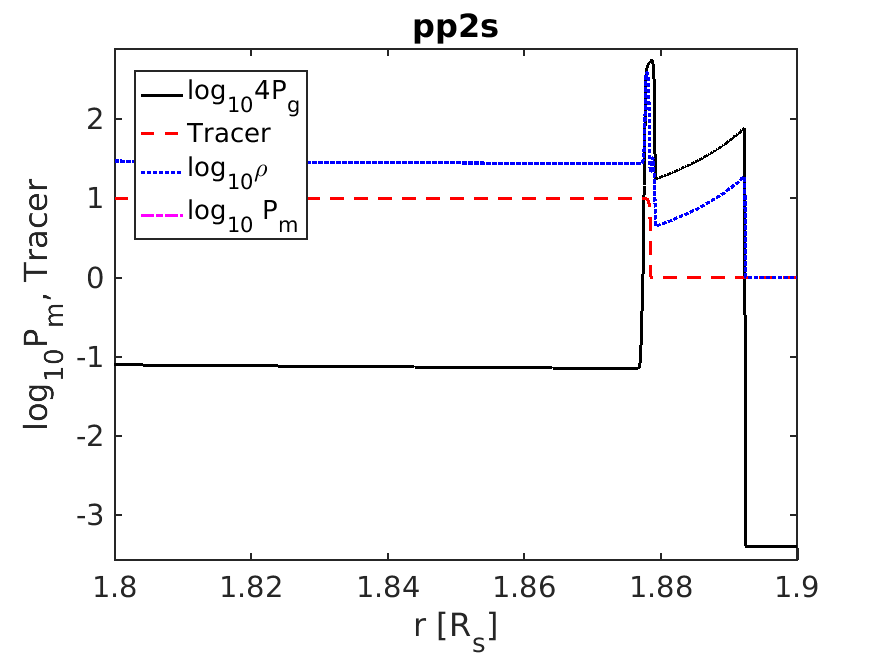}
\caption{Gas pressure (thick solid lines), density (dotted line)  and tracer 
(dashed line) as functions of   radius at the moment $t=1.9\; [r_s/c]$.}
\label{fig:Prrho}
\end{figure*}

\begin{figure*}
\includegraphics[width=0.48\linewidth,angle=-0]{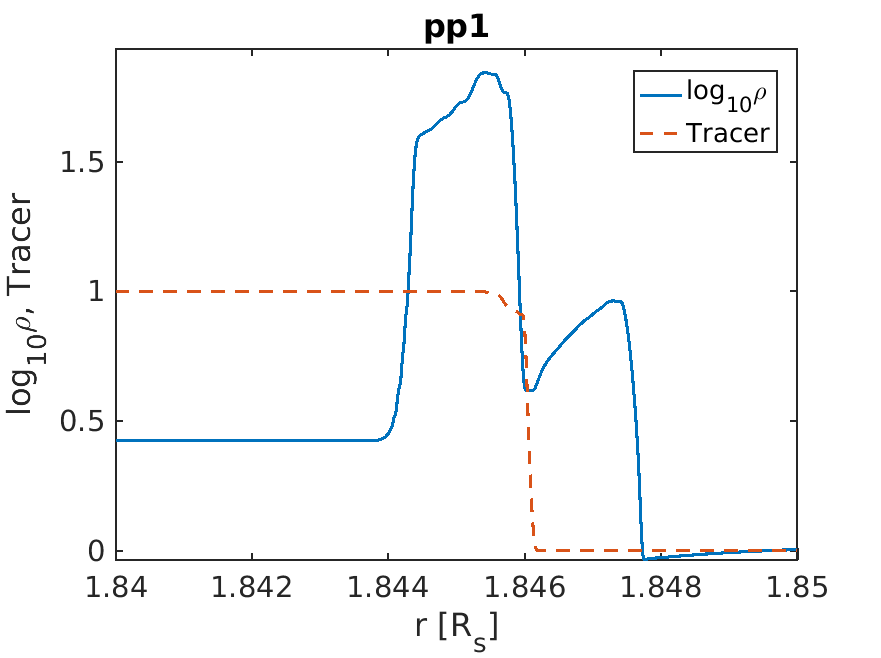}
\includegraphics[width=0.48\linewidth,angle=-0]{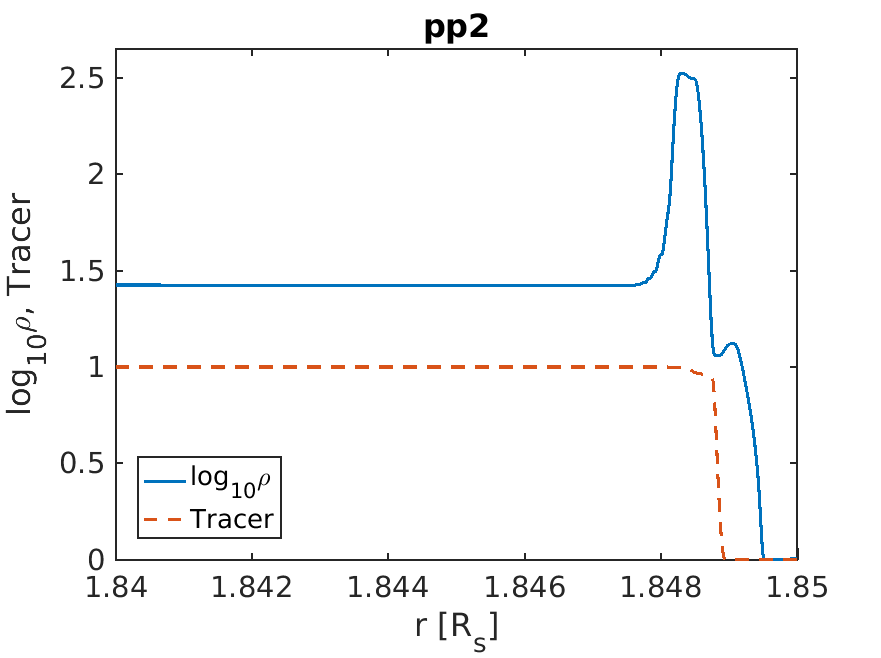}
\caption{Zoom-in to the region close to the CD: Density  (solid line)  and tracer (dashed line)  as functions of  radius at the moment $t=1.9\; [r_s/c]$.}
\label{fig:rhozoom}
\end{figure*}

As we can see on the Figure~(\ref{fig:Prrho}) from pm2  to pp2 model, with increasing   wind power,  the Lorentz factor of FS and RS are also increase while the  
distance between these shocks  becomes smaller,  where positions of the shocks are  indicated by jumps of pressure;   jump of density at constant pressure identifies the CD. 
Shift of the wind injection radius (compare models $pm2$ and $pm2s$ or $pp2$ and $pp2s$) do not 
change structure of the solution significantly. Change of injection radius shift position of shocked wind structure as a whole.  
High resolution of our setup allows to resolve structures of density distribution on the radial scale
$\sim 10^{-4}\; r_{\rm s}$ (see Figure~\ref{fig:rhozoom}).

\subsection{Magnetized cases}
\label{sec:a2}

\begin{figure*}
\includegraphics[width=0.48\linewidth,angle=-0]{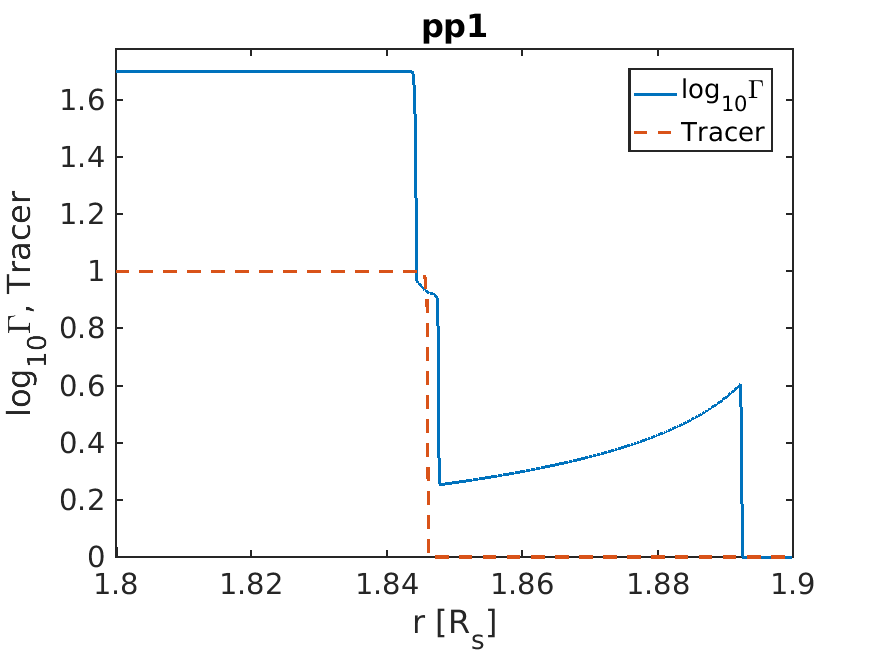}
\includegraphics[width=0.48\linewidth,angle=-0]{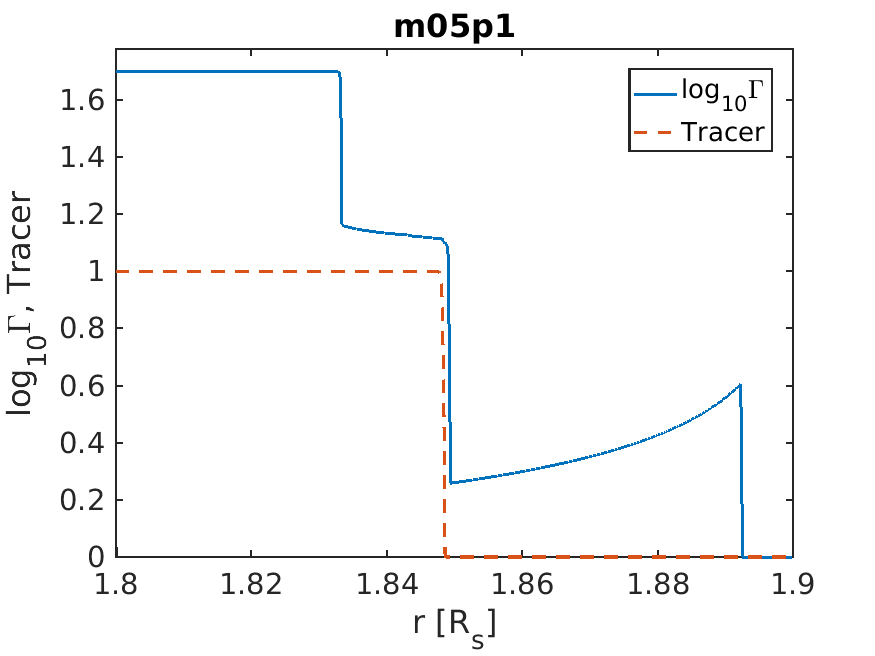}\\
\includegraphics[width=0.48\linewidth,angle=-0]{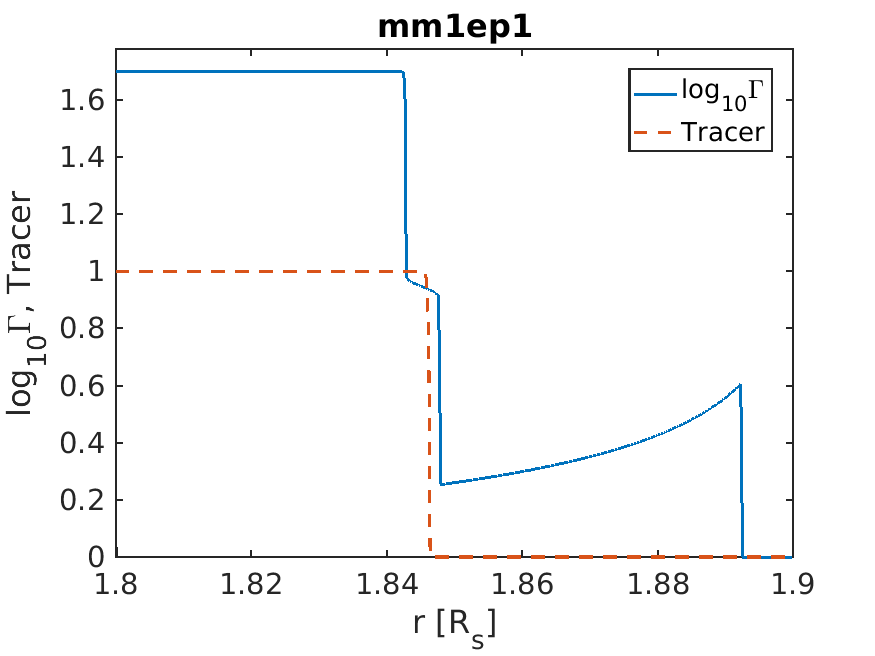}
\includegraphics[width=0.48\linewidth,angle=-0]{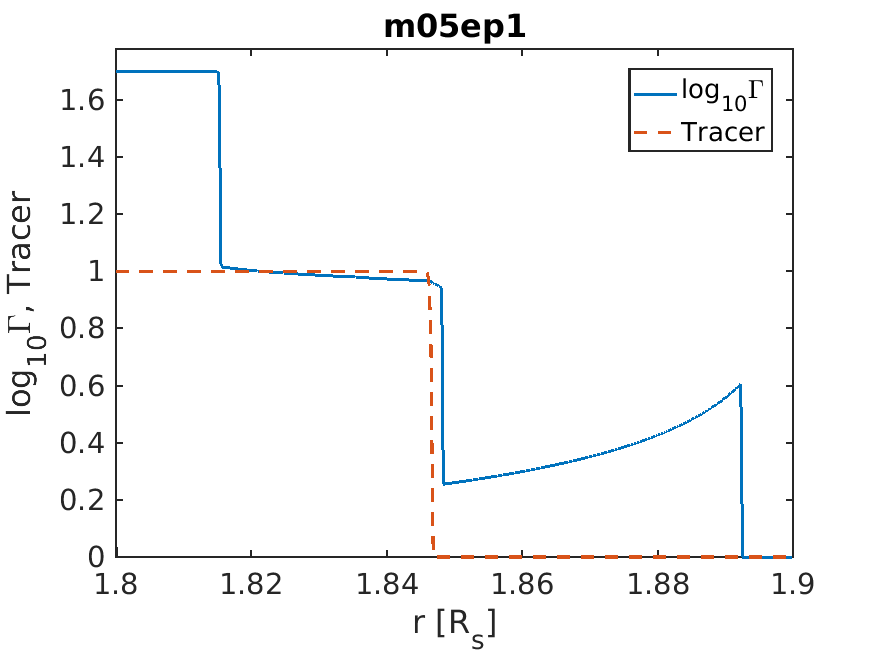}
\caption{Lorentz factor and tracer distribution as functions of 
radius at the moment $t=1.9$ for models with different magnetization.}
\label{fig:lorTr_mhd}
\end{figure*}

\begin{figure*}
\includegraphics[width=0.48\linewidth,angle=-0]{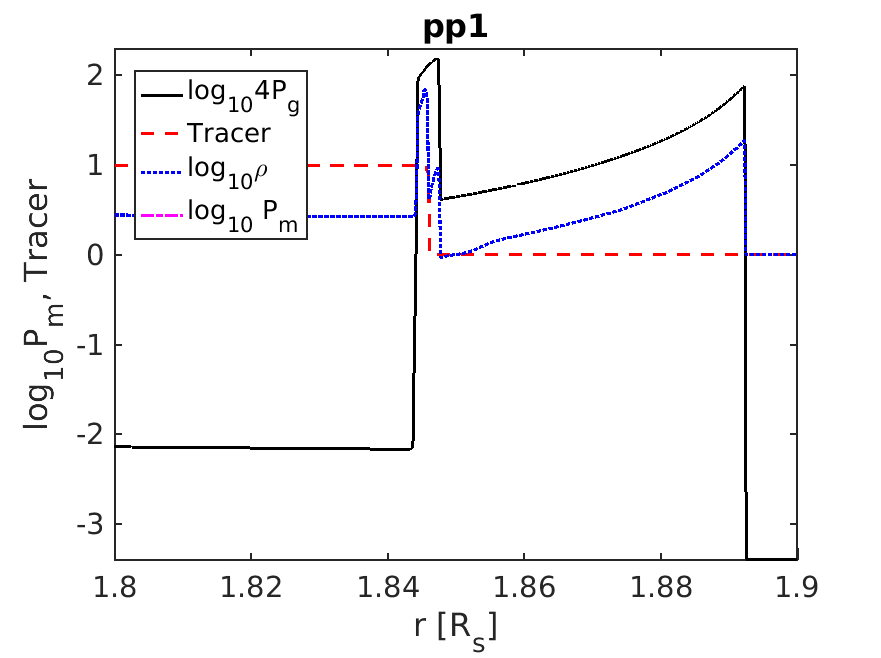}
\includegraphics[width=0.48\linewidth,angle=-0]{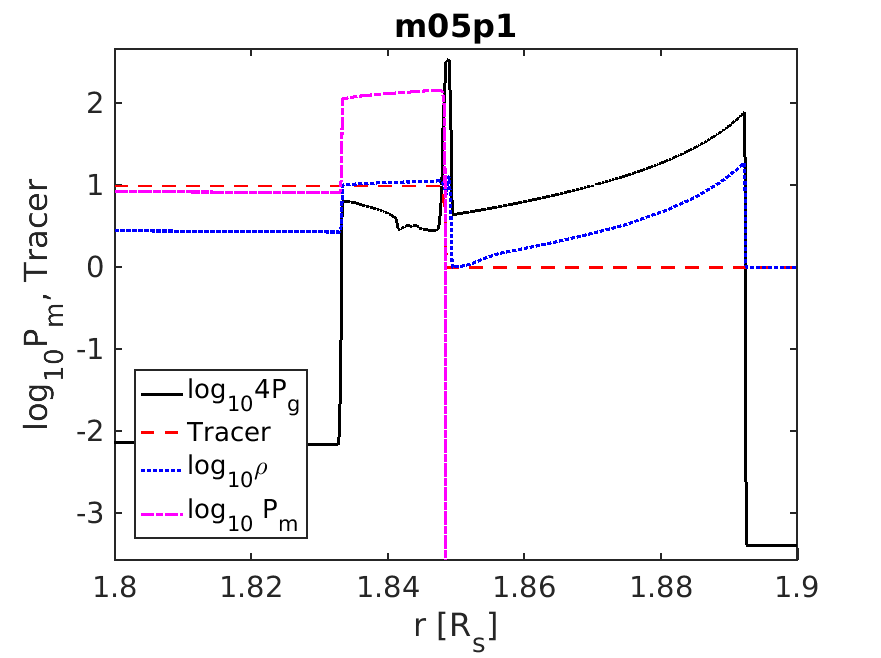}\\
\includegraphics[width=0.48\linewidth,angle=-0]{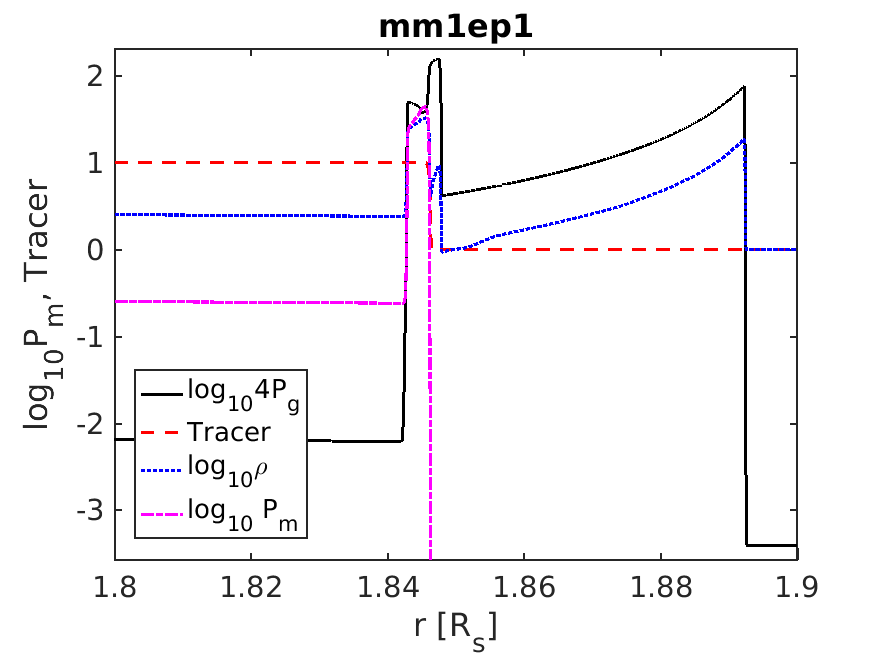}
\includegraphics[width=0.48\linewidth,angle=-0]{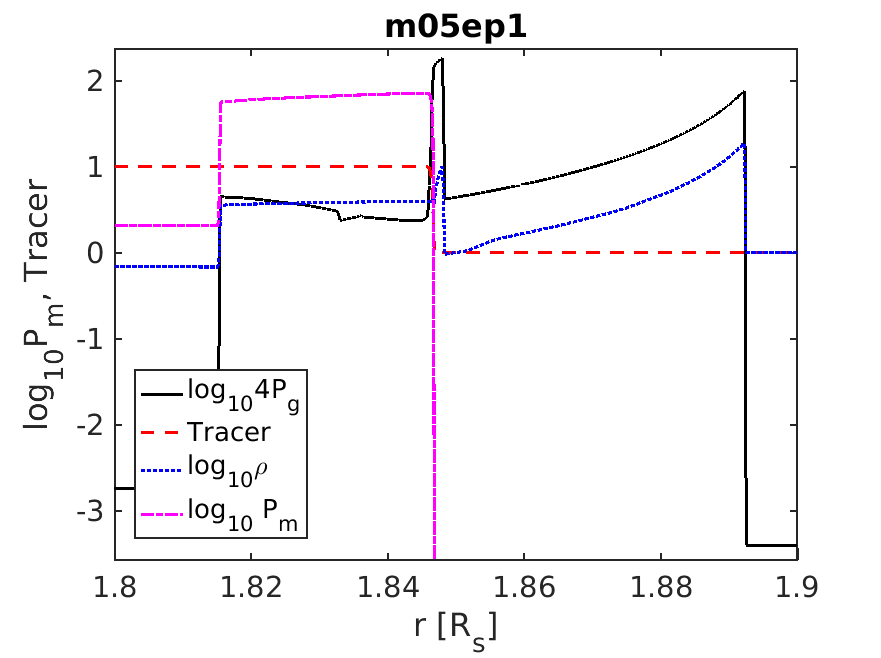}
\caption{Gas pressure (thick solid line), density (dotted line)  and tracer 
(dashed line) as functions of  radius at the moment $t=1.9$ for models with different magnetization.}
\label{fig:Prrho_mhd}
\end{figure*}

\begin{figure}
\includegraphics[width=0.48\linewidth,angle=-0]{pp1_rs_plato_rho_Tr_zoom_192000.png}
\includegraphics[width=0.48\linewidth,angle=-0]{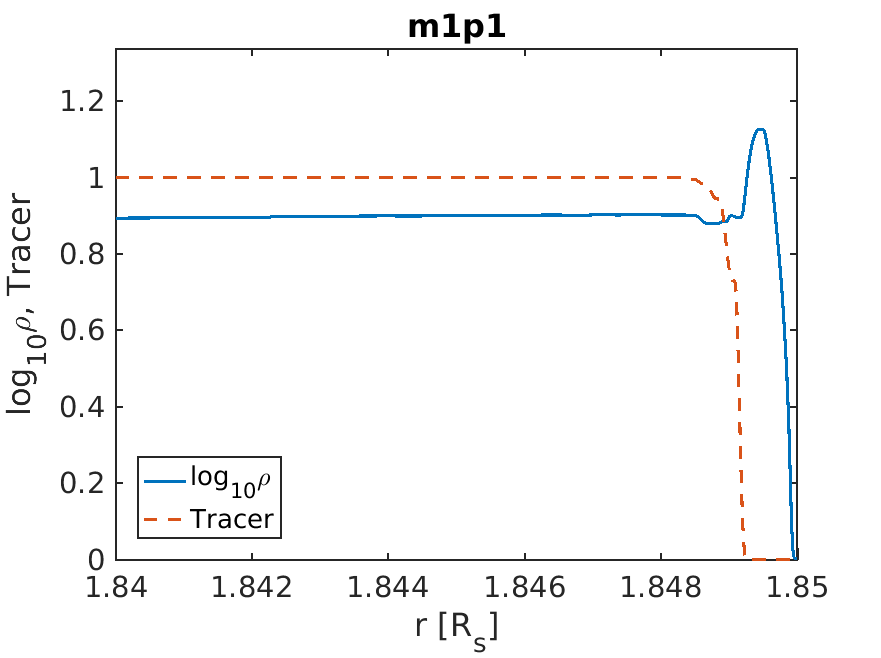}\\
\includegraphics[width=0.48\linewidth,angle=-0]{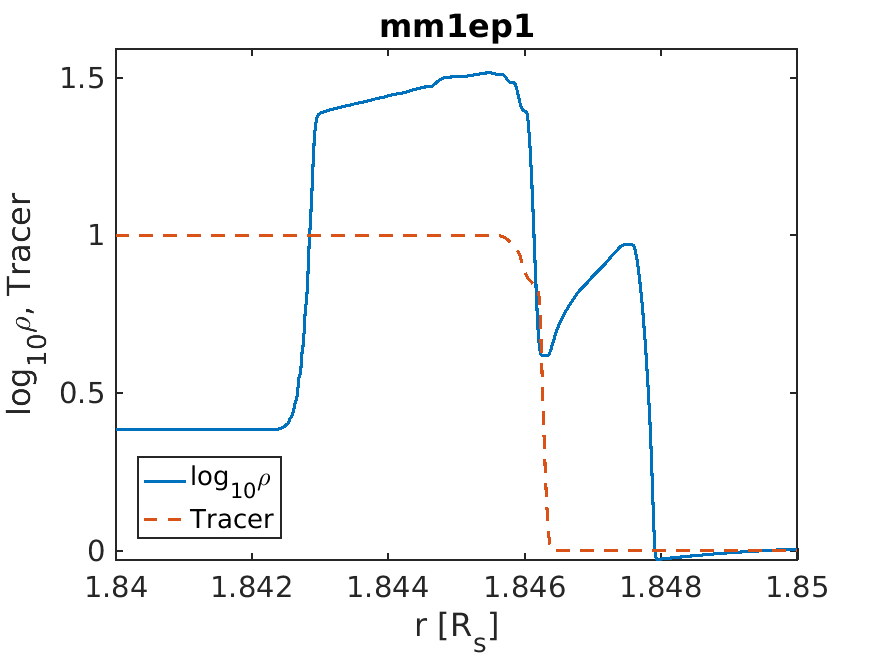}
\includegraphics[width=0.48\linewidth,angle=-0]{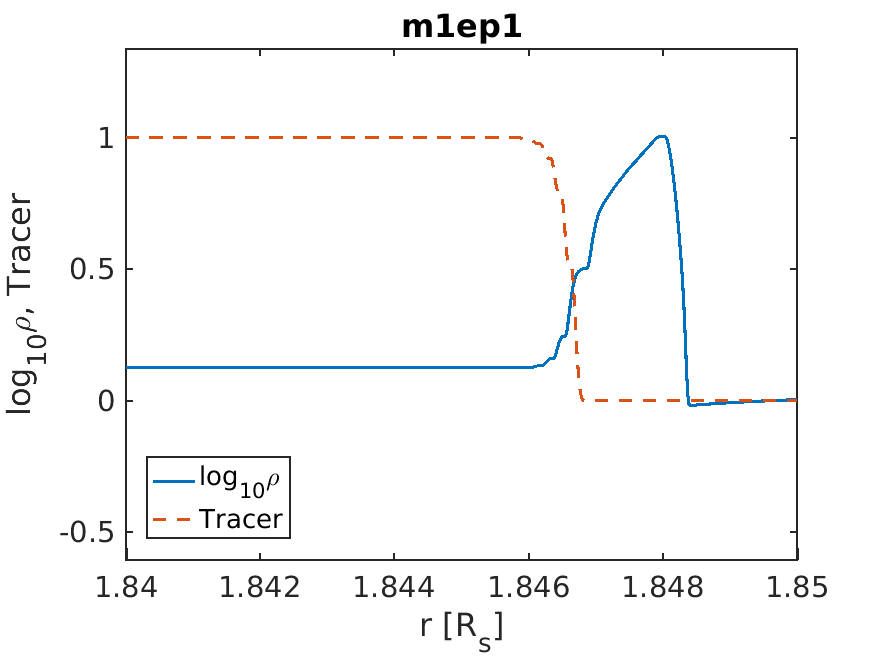}
\caption{Zoom-in to the regions near the CD. Density (solid line)  and tracer (dashed line) as functions of   radius at the moment $t=1.9$ 
for cases with different magnetization.}
\label{fig:rhozoom_mhd}
\end{figure}

Figures~(\ref{fig:lorTr_mhd}), (\ref{fig:Prrho_mhd}) and   \ref{fig:rhozoom_mhd}  demonstrate  weak dependence of density profile of double shocked matter if the total energy of the wind is preserved. On the other hand, 
  if we are preserving hydrodynamic energy flux in the wind   and increases its magnetization, due to increasing of 
  the total power of wind double shocked matter suffer stronger compression and layer double shocked matter became thinner. On other hand increase of magnetization 
  decrease compression ratio
of the shocked wind.

\bibliographystyle{apj}

\bibliography{BibTex,rs_plateau,reference}    


\appendix

\end{document}